\documentclass[12pt,a4paper]{article} 
\usepackage{graphicx}
\usepackage{amsmath}
\usepackage{amssymb}
\usepackage{mathtools}
\usepackage{dsfont}
\usepackage{slashed}
\usepackage{siunitx}
\usepackage{bbold}
\usepackage{color}
\usepackage[dvipsnames]{xcolor}
\usepackage[final]{changes} %
\usepackage[backend=bibtex,style=ieee,citestyle=numeric-comp]{biblatex}
\usepackage{hyperref}
\usepackage{cleveref}
\newcommand{\beq}{\begin{equation}}
\newcommand{\eeq}{\end{equation}}
\newcommand{\bea}{\begin{eqnarray}}
\newcommand{\eea}{\end{eqnarray}}

\newcommand{\One}{1\kern-4.5pt1}

\graphicspath{{figures/},{inkscape/}}
\newcommand{\im}{i}
\DeclareMathOperator{\real}{Re}
\DeclareMathOperator{\imag}{Im}
\DeclareMathOperator{\res}{Res}
\DeclareMathOperator{\asinh}{sinh^{-1}}
\DeclareMathOperator{\acosh}{cosh^{-1}}
\DeclareMathOperator{\ord}{\mathcal{O}}
\DeclareMathOperator{\tr}{Tr}
\newcommand{\ordnung}[1]{\ensuremath{\ord\left(#1\right)}}
\newcommand{\eto}[1]{\ensuremath{\mathrm{e}^{#1}}}
\newcommand{\md}{\ensuremath{\mathrm{d}}}

\makeatletter%
\begin{document}

\addtolength{\baselineskip}{0.20\baselineskip}

\hfill LTH/1315

\hfill October 2022

\begin{center}

\vspace{24pt}

{\LARGE {\bf Spectroscopy in the 2+1$d$ Thirring Model with $N=1$ Domain Wall Fermions} }

\end{center}

\vspace{18pt}

\centerline{{\bf  Simon Hands and Johann Ostmeyer}}

\vspace{20pt}

\centerline{Department of Mathematical Sciences, University of Liverpool,}
\centerline{Liverpool L69 3BX, U.K.}

\vspace{24pt}

\begin{center}  

{\bf Abstract}

\end{center}
We employ the domain wall fermion (DWF) formulation of the Thirring model on a
lattice in 2+1+1 dimensions and perform $N=1$ flavor Monte Carlo simulations.
At a critical interaction strength 
the model features a spontaneous $\mathrm{U}(2)\rightarrow\mathrm{U}(1)\otimes
\mathrm{U}(1)$ symmetry breaking;
we analyse the induced spin-0 mesons, both
Goldstone and non-Goldstone, as well as the correlator of the fermion
quasiparticles, in both resulting phases.
Crucially, we determine the anomalous dimension $\eta_\psi\approx 3$ at the
critical point, in stark
contrast with the Gross-Neveu model in 3$d$  and with results obtained with 
staggered fermions.  Our numerical simulations are complemented by an
analytical treatment of the free fermion correlator, which exhibits
large early-time artifacts due to branch cuts in the propagator stemming from
unbound interactions of the fermion with its heavy doublers.  These artifacts
are generalisable beyond the Thirring model, being an intrinsic property of DWF,
or more generally Ginsparg-Wilson fermions.\\


\vfill

\newpage

\section{Introduction}

The Thirring model is a quantum field theory of reducible (ie.\ 4-component) 
fermions interacting
via a current-current contact term, specified in three dimensional
continuum Euclidean
spacetime by the following Lagrangian:
\begin{equation}
        {\cal L}=\bar\psi_i(\partial{\!\!\! /}\,+m)\psi_i
        +{g^2\over{2N}}(\bar\psi_i\gamma_\mu\psi_i)^2,
\label{eq:thirring}
\end{equation}
with $i=1,\ldots,N$ indexing flavor degrees of freedom.
While (\ref{eq:thirring})  can be used to model electron dynamics in layered systems found in
condensed matter physics, it is theoretically interesting in its own right due
to its potential for exhibiting a UV-stable renormalisation group fixed point
where a strongly-interacting continuum quantum field theory may be defined. 
Since there is no small parameter in play, large anomalous scaling dimensions
are anticipated, so that the resulting theory will almost certainly lie
in a new universality class characterised by non-canonical critical exponents, a
scenario referred to as a Quantum Critical Point (QCP). This possibility may be
explored by several means; here we continue a programme of lattice field theory
simulations in which the QCP is identified in the $m\to0$ limit with a
transition in which the formation of a bilinear condensate $\langle\bar\psi\psi\rangle$
spontaneously breaks the model's global U(2$N$) symmetry leading to the dynamical
generation of a fermion mass. Further background can be found in recent
reviews~\cite{Hands:2021eyc,Wipf:2022hqd}.
 
The U(2$N$) symmetry of (\ref{eq:thirring}) is explicitly broken to
U($N$)$\otimes$U($N$) in the 
presence of a fermion mass $m\not=0$. Neither Wilson\added{ (with no
symmetry protecting against gap formation)} nor staggered \added{(where the
breaking is instead U($N)\otimes$U($N)\to$U($N$))} lattice
fermion formulations faithfully represent this pattern of breaking symmetry, which is problematic
when the QCP dynamics are strong and there is no means to control the recovery
of symmetry analytically. Our approach utilises domain wall fermions (DWF); on a
finite system with domain walls separated by $L_s$ in a direction $x_3$, there
is accumulated evidence both analytically and numerically that U(2$N$) is
recovered in the limit
$L_s\to\infty$~\cite{Hands:2015qha,Hands:2015dyp,Hands:2016foa,Hands:2020itv}.
Studies of the Thirring model with $N=1$ have revealed evidence for a QCP
described by an empirical equation of state for the order parameter
$\langle\bar\psi\psi(m,g)\rangle$ corresponding to critical exponents with
non-mean field values~\cite{Hands:2020itv,Hands:2021smr}, and distinct from
those obtained from simulations of the model formulated with staggered lattice
fermions~\cite{DelDebbio:1997dv}. Moreover, simulations with $N=2$ have failed
to identify a condensate for $m\to0$, consistent with a critical flavor number
$1<N_c<2$, with $N\leq N_c$ needed for the QCP's existence~\cite{Hands:2018vrd},
and again in disparity with the $N_c\approx7$ observed for staggered
fermions~\cite{Christofi:2007ye}\footnote{\added{It has been proposed that the staggered model 
captures a continuum model based on K\"ahler-Dirac fermions, which has a
distinct global symmetry~\cite{Hands:2021mrg}.}}.

\added{In this paper we turn our attention to
two-point functions},
\deleted{(following exploratory but inconclusive work presented in
\cite{Hands:2016foa,Hands:2018vrd})} 
studying both fermion -- antifermion ``meson'' bound states
in the spin-0 channel, and also the propagating fermion ``quasiparticle'' in the
spin-${1\over2}$ channel.  The calculations employ orthodox lattice field
theory techniques, and for a massive theory, which we can ensure by setting
$m\not=0$, yield information on the
particle spectrum. Since the quasiparticle propagator is not gauge-invariant in
a gauge theory, to our knowledge this is the first time elementary fermion
excitations have been studied using DWF. In  principle it will enable us to 
distinguish broken from symmetric phases via dynamical fermion mass
generation and the appearance of Goldstone bosons whose mass has a
characteristic dependence on $m$ in the former case. However, the DWF setup also
enables a study of the massless $m=0$ limit, in which case exactly at the critical
point $g=g_c$ all correlations are expected to decay algebraically, with a power
sensitive to the critical dynamics. Study of the quasiparticle propagator 
now furnishes information on a critical exponent $\eta_\psi$, defined via 
$\langle\psi(x)\bar\psi(0)\rangle\sim x^{-(2+\eta_\psi)}$, an important
characteristic of the QCP {\em not\/} accessed via the equation of
state, which focusses on the scalar order parameter field. 

The rest of the paper is organised as follows. In Sec.~\ref{sec:form} we recall
the lattice formulation of the Thirring model (\ref{eq:thirring}) with DWF,  and
define the correlation functions to be calculated in terms of fields
$\Psi,\bar\Psi$ defined on a 2+1+1$d$ lattice. Since the study of elementary
fermion propagators is new, analytic insight is welcome\added{; in
Sec.~\ref{sec:free} we have
attempted to collect results for the free fermion propagator
which will inform studies in both this and future
work, including} 
\deleted{Sec.~\ref{sec:free} presents results for the free
fermion
propagator, along with} a simple analytic model which to good accuracy reproduces a numerically
significant artifact resulting from a branch cut in the exact form
(\ref{eq:exact_mom_prop_h}) below. It is demonstrated that finite-$L_s$ artifacts are
reduced if instead of $m\bar\psi\psi$ the U($2N$)-equivalent mass term
$im\bar\psi\gamma_3\psi$ is used, corroborating earlier studies of the condensate
$\langle\bar\psi\psi\rangle$~\cite{Hands:2015qha, Hands:2015dyp,Hands:2016foa}.
Dependence on the domain wall height $M$ is also studied. Sec.~\ref{sec:results}
presents spectrum results from numerical simulations on a $16^2\times48\times L_s$ system with
varying $m,g$ and the domain wall separations
$L_s=64,80$ used in the most recent equation-of-state
study~\cite{Hands:2021smr}. 
Our fitting procedure is described in detail; 
we study 3 distinct spin-0 mesons including both
Goldstone and non-Goldstone channels (a non-Goldstone requiring
the evaluation of disconnected diagrams is omitted for now) and the fermion
quasiparticle. In Sec.~\ref{sec:conformal} we turn our attention to the fermion
channel with $m$ set to zero. First we develop continuum models for the quasiparticle propagator
at a QCP, ie.\ with anomalous dimension $\eta_\psi\not=0$, showing that in
general a UV regularisation is required. We also present an {\em
Ansatz} for how the propagator might be  modified away from the QCP in the symmetric
phase, ie.\ with a finite
correlation length $\mu^{-1}$ where $\mu$ is {\em not\/} a pole mass. 
Finally we present numerical results taken at 5
different couplings in the symmetric phase including one very close to the QCP
deduced from equation of state studies~\cite{Hands:2021smr}, obtaining an
estimate $\eta_\psi\approx3$ at the critical point. Sec.~\ref{sec:disco}
discusses our results, with some technical details postponed to 
the \Cref{sec:mats,sec:dw_height_separation,sec:IR}.

\section{Lattice Formulation and Methodology}
\label{sec:form}

The lattice model studied here is the ``bulk'' variant of the 
formulation first set out in \cite{Hands:2016foa}, employing domain wall
fermions (DWF) in 2+1+1$d$:
\begin{equation}
S=S_{\rm kin}+S_{\rm int}+S_{\rm
aux}=\sum_{x,y}\sum_{s,s^\prime}\bar\Psi(x,s){\cal
M}_{x,s;y,s^\prime}\Psi(y,s^\prime)+S_{\rm aux}.
\label{eq:lattice_thirring}
\end{equation}
Here $\Psi,\bar\Psi$ are 4-spinors defined on a hypercubic lattice with 2+1$d$
indices $x$ and an index $s$ labelling the ``third'' direction $x_3$, taking values
$s=1,\ldots,L_s$. Free fermions are described by the kinetic operator
\begin{equation}
{\cal M}_0 =\delta_{s,s^\prime}D_{Wx,y}+\delta_{x,y}D_{3s,s^\prime}+mS_{m3};
\label{eq:free-fermion}
\end{equation}
$D_W$ is the 2+1$d$ Wilson operator with domain wall height $M$:
\begin{equation}
D_W(M)_{x,y}=-{1\over2}\sum_{\mu=0,1,2}[(1-\gamma_\mu)\delta_{x+\hat\mu,y}+(1+\gamma_\mu)\delta_{x-\hat\mu,y}]+(3-M)\delta_{x,y}.
\end{equation}
Throughout this work we use $M=1$. Hopping along $x_3$ is governed by
\begin{equation}
D_3(L_s)_{s,s^\prime}=-[P_-\delta_{s+1,s^\prime}(1-\delta_{s,L_s})+P_+\delta_{s-1,s^\prime}(1-\delta_{s,1})]+\delta_{s,s^\prime}.
\label{eq:Shamir}
\end{equation}
The factors $(1-\delta_{s,1/L_s})$ implement open boundary conditions at
domain walls located at $s=1,L_s$, while the projectors
$P_\pm={1\over2}(1\pm\gamma_3)$ also appear in the definition of the target
physical fermion degrees of freedom $\psi,\bar\psi$ defined on the walls:
\begin{equation}
\psi(x)\equiv P_-\Psi(x,1)+P_+\Psi(x,L_s);\;\;\;
\bar\psi(x)\equiv \bar\Psi(x,L_s)P_-+\bar\Psi(x,1)P_+.
\label{eq:target}
\end{equation}
Finally, the mass term is defined in terms of fields on the walls via
\begin{equation}
mS_{m3}
=im\sum_x\bar\psi(x)\gamma_3\psi(x).\label{eq:mass_3}
\end{equation}
This form of the mass term yields 
superior convergence to the U(2)-symmetric limit anticipated as $L_s\to\infty$ over
the conventional $m\bar\psi\psi$
~\cite{Hands:2015qha,Hands:2016foa}.

The interaction term is between a fermion current and a real non-compact
vector field $A_\mu$ defined on the links of the spacetime lattice:
\begin{eqnarray}
S_{\rm int}&=&{i\over2}\sum_{x,\mu,s}A_\mu(x)\left[
\bar\Psi(x,s)(-1+\gamma_\mu)\Psi(x+\hat\mu,s)+\bar\Psi(x+\hat\mu,s)(1+\gamma_\mu)\Psi(x,s)\right]\nonumber\\
&\equiv&\sum_{x,\mu}A_\mu(x){\cal J}_\mu(x)
\end{eqnarray}
Integration over the auxiliary field specified by the Gaussian action
\begin{equation}
S_{\rm aux}={1\over{2g^2}}\sum_{x,\mu}A_\mu^2(x),
\end{equation}
results in a four-fermion contact interaction $-{g^2\over2}{\cal J}_\mu{\cal J}_\mu$ between
conserved non-local currents 
\begin{equation}
{\cal J_\mu}(x) = \sum_{s=1}^{L_s} j_\mu(x,s);\;\;\;
\Delta^-_\mu{\cal J}_\mu(x)=\sum_s\Delta_\mu^-j_\mu(x,s)=0.
\end{equation}
with the local current ($\nu\in\{\mu,3\}$)
\begin{equation}
j_\nu(x,s)={i\over2}\left[\bar\Psi(x,s)(\gamma_\nu-1)\Psi(x+\hat
\nu,s)+\bar\Psi(x+\hat \nu,s)
(\gamma_\nu+1)\Psi(x,s)\right]
\end{equation}
obeying a 2+1$d$ continuity equation\footnote{If the U(2)-equivalent mass term
$im\bar\psi\gamma_5\psi$ is used~\cite{Hands:2015qha}, there are no  terms proportional to
$m$ in (\ref{eq:continuity}).} (Cf.~\cite{Furman:1994ky}):
\begin{equation}
\Delta_\mu^-j_\mu(x,s)=\begin{cases}
-j_3(x,1)-m\bar\psi(x)\psi(x)
& s=1\\
-\Delta_3^-j_3(x,s) & 1<s<L_s\\
+j_3(x,L_s-1)+m\bar\psi(x)\psi(x)
& s=L_s.
\end{cases}
\label{eq:continuity}
\end{equation}

The fermion matrix ${\cal M}[A_\mu]$ superficially resembles that of an abelian
gauge theory, with link fields $e^{\pm iA_\mu}$ replaced by non-unitary links
$(1\pm iA_\mu)$. At strong coupling this non-unitary nature presents
challenges, both in inverting ${\cal M}$ and in the recovery of U(2) as
$L_s\to\infty$~\cite{Hands:2020itv}. In practice $N=1$ dynamics are simulated 
with an RHMC algorithm based on the positive measure $\mbox{det}({\cal
M}^\dagger{\cal M})^{1\over2}$; details can be found in
\cite{Hands:2018vrd,Hands:2020itv} and the simulation code is available at
\cite{code}.

Quasiparticle and meson propagators are calculated in terms of the 2+1+1$d$
propagator $S(m;x,s;y,s^\prime)=\langle\Psi(x,s)\bar\Psi(y,s^\prime)\rangle$, which
obeys two useful identities~\cite{Hands:2015qha}:
\begin{eqnarray}
\gamma_5S(m;x,s;y,s^\prime)\gamma_5&=&S^\dagger(m;y,s^\prime;x,s);\label{eq:gamma5}\\
\gamma_3S(m;x,s;y,s^\prime)\gamma_3&=&S^\dagger(-m;y,\bar s^\prime;x,\bar
s),\label{eq:gamma3}
\end{eqnarray}
with $\bar s\equiv L_s-s+1$. Meson propagators in spin-0 channels are then
defined using local bilinear sources via 
\begin{equation}
C_\Gamma(x)=\langle
\bar\psi(0)\Gamma\psi(0)\bar\psi(x)\Gamma\psi(x)\rangle;\;\;\;
\Gamma\in\{\gamma_3,\gamma_5,\One,\gamma_3\gamma_5\}.
\end{equation}
Using the definition (\ref{eq:target}) and relations (\ref{eq:gamma5},\ref{eq:gamma3}) 
they can all be expressed in
terms of the primitive correlators~\cite{Hands:2015qha,Hands:2018vrd}
\begin{eqnarray}
C^{--}(x)&=&\mbox{tr}[S(m;0,1;x,L_s)P_-S^\dagger(m;0,1,x,L_s)P_-];\nonumber\\
C^{+-}(x)&=&\mbox{tr}[S(m;0,1;x,1)P_+S^\dagger(m;0,1,x,1)P_-];\nonumber\\
\tilde
C^{--}(x)&=&\mbox{tr}[S(m;0,1;x,L_s)P_-S^\dagger(-m;0,1,x,L_s)P_-];\nonumber\\
\tilde C^{+-}(x)&=&\mbox{tr}[S(m;0,1;x,1)P_+S^\dagger(-m;0,1,x,L_s)P_-],
\end{eqnarray}
requiring two inversions of ${\cal M}$ for each source location on the 
$s=1$ wall.
The resulting expressions are
\begin{alignat}{3}
C_{\gamma_5}(x)&\equiv C_{\rm G^-}(x)&=\vert C^{--}(x)+C^{+-}(x)\vert;\label{eq:meson_corr_g5}\\
C_{\One}(x)&\equiv C_{\rm G^+}(x)&=\vert\tilde C^{--}(x)-\tilde C^{+-}(x)\vert;\label{eq:meson_corr_id}\\
C_{\gamma_3}(x)&\equiv C_{\rm NG^+}(x)&=\vert\tilde
C^{--}(x)+\tilde C^{+-}(x)\vert;\label{eq:meson_corr_g3}\\
C_{\gamma_3\gamma_5}(x)&\equiv C_{\rm NG^-}(x)&=\vert
C^{--}(x)-C^{+-}(x)\vert.\label{eq:meson_corr_g3g5}
\end{alignat}
The channel subscripts denote whether the meson is Goldstone or non-Goldstone,
based on an anticipated U(2)$\to$U(1)$\otimes$U(1) symmetry breaking induced by
a symmetry-breaking mass term $im\bar\psi\gamma_3\psi$. Parity $\pm$ assignments
follow the definition 
\begin{equation}
\psi(x)\overset{\cal P}\mapsto\gamma_3\psi(-x);\;\;\;
\bar\psi(x)\overset{\cal P}\mapsto\bar\psi(-x)\gamma_3\,,
\end{equation}
chosen to leave this mass term invariant. Note that in the case of
symmetry breaking the NG$^+$ channel also has a significant contribution of the
opposite sign from disconnected fermion line diagrams, which we do not attempt
to calculate.

In the spin-${1\over2}$ sector the timeslice propagator for free
fields is
\begin{equation}
S_f(x_0)=\sum_{\vec x}\langle\psi(0)\bar\psi(x)\rangle\sim
\int{dp_0\over2\pi}{e^{ip_0x_0}\over{ip_0\gamma_0+im\gamma_3}}
={{-i\gamma_3\pm\gamma_0}\over2}e^{-m\vert x_0\vert},
\end{equation}
where $\pm$ denotes the sign of the temporal displacement $x_0$. In terms of 
2+1+1$d$ propagators this motivates the measurements
\begin{eqnarray}
S_0(x_0)={1\over4}\mbox{tr}\gamma_0 S_f(x_0)\!\!\!&=&\!\!\!{1\over4}
\sum_{\vec x}\mbox{tr}\gamma_0[P_-S(m;0,1;x,1)+P_+S(m;0,L_s;x,L_s)];\nonumber\\
S_3(x_0)={i\over4}\mbox{tr}\gamma_3 S_f(x_0)\!\!\!&=&\!\!\!{i\over4}
\sum_{\vec x}\mbox{tr}[-P_-S(m;0,1;x,L_s)+P_+S(m;0,L_s;x,1)].\nonumber\\
\label{eq:quasiprop}
\end{eqnarray}
For enhanced sampling expressions (\ref{eq:quasiprop}) are evaluated using a
wall source; since there is no need for gauge-fixing, this presents no additional
complications. 
We sampled every 5th trajectory, using
sources located at 5 different timeslices, each
requiring separate inversions on two distinct Dirac-indexed sources to evaluate
(\ref{eq:quasiprop}).
This was found to yield substantially improved results compared to earlier studies 
employing smeared sources~\cite{Hands:2016foa}.

\section{Free Fermion Correlator}
\label{sec:free}

In this section we collect together some analytic results and approximations for the free fermion
correlator using DWF, with the goal of understanding \replaced{the discretisation effects introduced by the lattice as well as the influence of domain wall height $M$ and separation $L_s$}{how well the
continuum form is recovered as domain wall height $M$ and separation $L_s$ are
varied}. \added{All the considered formulations have the correct continuum limit, but they approach it differently swiftly which can be crucial for simulations with limited computational resources.} This will inform the numerical study of the fermion propagator in the interacting
theory to be presented in Secs.~\ref{sec:results},\ref{sec:conformal}.

We explicitly distinguish between the two mass operators $\im \gamma_3 m_3$ as in equation~\eqref{eq:mass_3} and the conventional hermitian $m_h$ in this section.
Since we are interested in analytic results here, we derive the free fermion propagator in terms of operators rather than the measured quantities $S(m;x,s;y,s^\prime)$. Furthermore we work in momentum space. Following Ref.~\cite{Hands:2016foa}, we obtain the expressions (formally equivalent to those in equation~\eqref{eq:quasiprop})
	\begin{align}
		C(p;1,1) &= \tr\left[\gamma_0P_-D^\dagger(p;1,s)G(p;s,1)\right]\,,\label{eq:corr_same_wall}\\
		C(p;1,L_s) &= \begin{cases}
			\tr\left[P_-D^\dagger(p;1,s)G(p;s,L_s)\right]\,, & m=m_h\\
			\tr\left[-\im\gamma_3 P_-D^\dagger(p;1,s)G(p;s,L_s)\right]\,, & m=\im\gamma_3 m_3
		\end{cases}\label{eq:corr_diff_walls}
	\end{align}
	where the Wilson operator $D$ (corresponding to  $\cal M_0$ in
(\ref{eq:free-fermion})) and the Green function $G$ are defined by
	\begin{align}
		D^\dagger(p;1,s) &= \theta(s-1)\theta(L_s-s)\left[-P_+\delta_{s,2}+(b-\im\slashed{\bar p})\delta_{s,1}+mP_-\delta_{s,L_s}\right]\,,\\
		\begin{split}
			G(p;s,s') &= \left(P_+A_++P_-A_-\right)\eto{-\alpha(s+s'-2)} + \left(P_+A_-+P_-A_+\right)\eto{-\alpha(2L_s-s-s')}\\
			&\quad + B\eto{-\alpha|s-s'|} + A_m\left(\eto{-\alpha(L_s-s+s'-1)}+\eto{-\alpha(L_s+s-s'-1)}\right)
		\end{split}
	\end{align}
	respectively, with the auxiliary variables $\alpha, \bar p, b, A_\pm, A_m$, and $B$ listed in \Cref{sec:aux_vars}.
	
	For zero spatial momentum, the explicit evaluation of the traces in~\eqref{eq:corr_same_wall} and~\eqref{eq:corr_diff_walls} yields the form
	\begin{align}
		C(p=(p_0,0,0)) &= C((p_0,0,0);1,1)+C((p_0,0,0);1,L_s)\\
		&= \frac{2 \im m \sin (p_0)+2 \cos (p_0)+\sqrt{5-4 \cos (p_0)}-1}{\im \left(m^2+1\right) \sin (p_0)+2 m \cos (p_0)-m}\label{eq:exact_mom_prop_h}
	\end{align}
	in the large $L_s$ limit and setting the domain wall height $M=1$ immediately. 
The exact form for finite $L_s$ can be found in \Cref{sec:dw_height_separation}.

	We obtain the same expression for both mass terms, $\im \gamma_3 m_3$ and $m_h$,
	in the $L_s\rightarrow\infty$ limit as expected. The convergence in the $m_3$
case is significantly faster, however, as we will discuss later on.

	\subsection{The free propagator in momentum and real space}
\label{sec:free_simple}
 Often the
physical intuition obtained from exact analytic
calculations is rather  limited. We will  therefore investigate an
approximation that captures all the important physics without
`having too many trees to see the forest'.
	
	We start out with the well known (up to a constant and irrelevant factor 2) free particle propagator in continuous
space\footnote{This works straightforwardly at zero momentum as then $p\equiv
p_0$ is scalar which is the relevant case for now, but it can also be extended
canonically to the vectorial version.} 
\begin{align} 
C_{\rm cont}(p) &= \frac{2}{m+\im
p}\,.  
\end{align} 
Going to a lattice, we have to substitute the momentum $p$
for the lattice momentum $\sin p$.
The simplest
realisation with the correct continuum limit is then
\begin{align} 
C_{\rm naive}(p) &= \frac{2}{m+\im \sin p}\,, 
\end{align} 
which, of course,
leads to the infamous doubling problem since the $\sin$-function has
zeros not only at integer multiples of $2\pi$, but also at multiples of $\pi$.
The domain wall approach essentially gets rid of this problem by lifting the
unphysical pole near $p=\pi$ (assuming $m\ll 1$). The exact
formula~\eqref{eq:exact_mom_prop_h} is a particular realisation of this
requirement, and so is 
\begin{align} 
C_0(p) &= \frac{1+n(m)\cos p}{m+\im \sin
p}\,,\quad n(m) \coloneqq \frac{1}{\sqrt{1+m^2}}\,. 
\end{align}

\begin{figure}[htp]
		\centering
		\includegraphics[width=.45\textwidth]{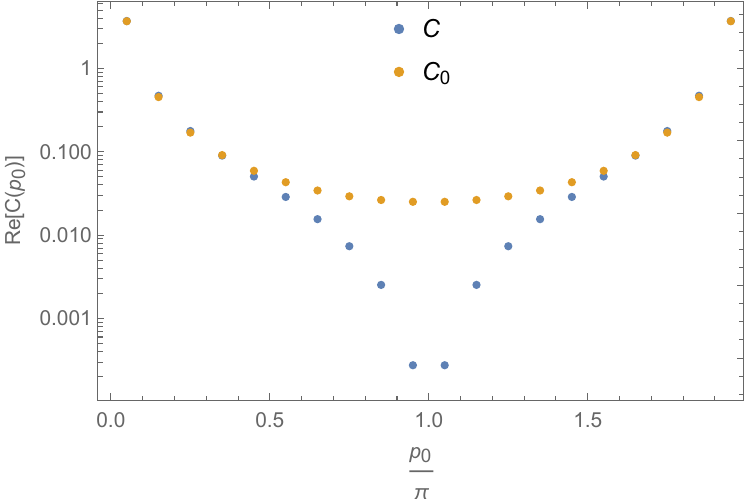}
		\includegraphics[width=.45\textwidth]{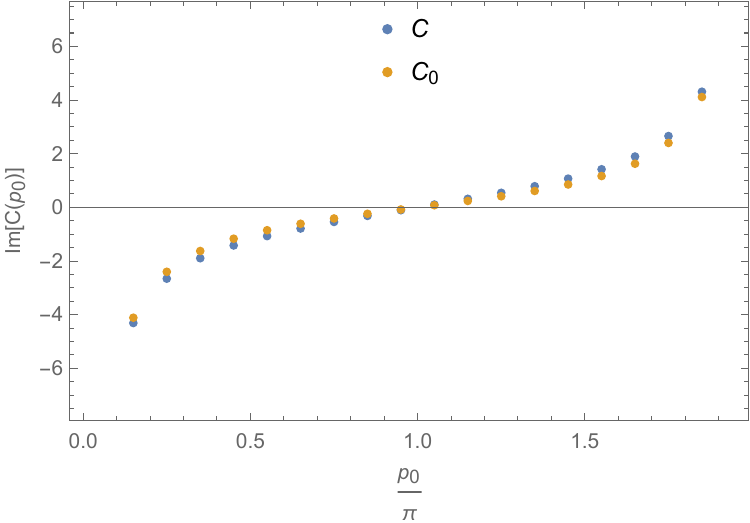}
		\caption{Exact and approximate free fermion propagators at zero spatial momentum in momentum space, $m=\num{0.05}$. Real part left, imaginary part right.}\label{fig:free_prop_mom_space}
	\end{figure}

	Figure~\ref{fig:free_prop_mom_space} shows that $C$ and $C_0$ are quite compatible, 
so we are going to use the simplistic version $C_0(p)$ in further analysis.

	We are interested in the propagator in (imaginary) time, still at zero spatial momentum, so we have to perform a Fourier transformation
	\begin{align}
	C_0(t)&=\frac{1}{L_t}\sum_{p_0}C_0(p)\eto{\im p_0 t}\,,
	\end{align}
where $p_0$ ranges over fermionic Matsubara modes.
	The full derivation of the exact form is provided in \Cref{sec:c0_derivation} and it yields
	\begin{align}
		C_0(t)
		&=\frac{\eto{-\tilde m t}}{\eto{-\tilde m L_t}+1}\frac{2}{\sqrt{1+m^2}}-\delta_{t0}\,,
	\end{align}
	where $\tilde m\coloneqq \asinh m$. In the zero temperature
($L_t\rightarrow\infty$) and continuum ($m\rightarrow0$, but $mt=\text{const.}$)
limits $C_0(t)$ approaches the expected form $\eto{-m t}$.

	\subsection{Leading order corrections}
	The propagator \added{$C_0$} derived in Sec.~\ref{sec:free_simple} captures the important
intermediate time $1\ll t \ll L_t$ features including some lattice artefacts and
the absence of the doubler. There are, however, more subtle but still
substantial discretisation effects not yet considered. When considering the
exact form $C(p)$ instead of the simplified $C_0(p)$, the most prominent
difference is that $C(p)$ has not only poles but also branch cuts due to the
$\sqrt{5-4\cos p}$ term. The square root stems from the quadratic nature
of $DD^\dagger$ solved for the Greens function~\cite{Hands:2016foa}, or more
generally from the quadratic (chirality breaking) term of order $\ordnung{a}$ in
the Ginsparg-Wilson equation~\cite{Ginsparg_Wilson}. While all the other
differences are analytic and can therefore feature only in higher orders of $m$,
this non-holomorphicity has an immediate impact on the contour integral and
therefore on $C(t)$. The existence of branch cuts in the DWF 
representation has been noted before~\cite{Gavai_2009}, but to the best of our
knowledge so far neither their origins nor their implications have been
investigated.
	
	Branch cuts indicate unbound many-particle
interactions~\cite{Peskin:1995ev}, in this case between the fermion and its
doubler. These interactions are very short ranged for heavy doublers and
decouple completely in the continuum limit. Put differently, the branch cuts can
only start at energies larger than the sum of fermion and doubler masses and
therefore vanish in the limit of infinitely heavy doublers. Here we see again
that DWF (or more generally Ginsparg-Wilson fermions) do not get rid of the
doublers in principle, but rather assign zero weight to the single particle doubler
poles.
	
\begin{figure}[htp] \centering
\includegraphics[width=.9\textwidth]{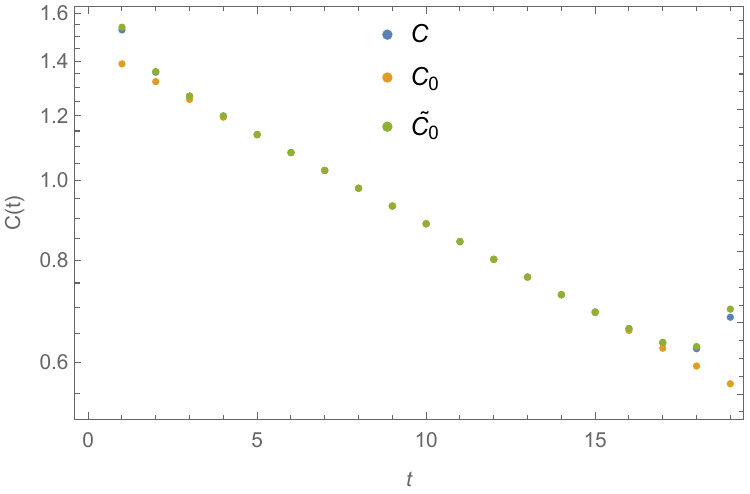} \caption{Exact and
approximate free fermion propagators at zero spatial momentum in real space,
$m=\num{0.05}$.}\label{fig:free_prop_real_space} \end{figure}

	Again, the details of the modified correlator's derivation can be found
in \Cref{sec:c0_tilde_derivation}.	We call the modified propagator that
incorporates both $C_0$ and the branch cuts 
\begin{align} \tilde C_0(t) &=
C_0(t) +
\sqrt{\frac3{4\pi}}\left(\frac{1}{m+\frac34}\frac{2^{t-L_t}}{\left(L_t-t+\frac34\right)^{3/2}}
+ \frac{1}{\frac34-m}\frac{2^{-t}}{\left(t+\frac34\right)^{3/2}}\right)
\end{align} 
and we show all three propagators as a function of $t$ in
figure~\ref{fig:free_prop_real_space}. As expected the unphysical contributions
vanish exponentially fast when $t$ and $L_t-t$ are large\added{, so that the
correct results are obtained in the continuum and zero temperature limits}. Careful zooming in
reveals some small differences between $C$ and $\tilde C_0$ at the edges of
the diagram, but the leading order features are described very well.
	
	\subsection{Influences of the Domain wall height and separation}
	Throughout this work we set the DW height $M=1$ and therefore do not go
into detail about its influence on the propagator here. A
short summary of how $M$ affects the propagator is provided in \Cref{sec:dw_height_separation}.
	
\begin{figure}[htp]
		\centering
		\includegraphics[width=.45\textwidth]{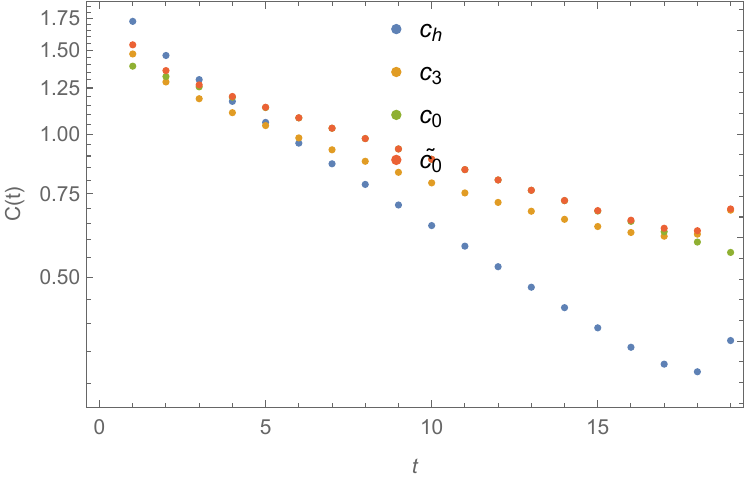}
		\includegraphics[width=.45\textwidth]{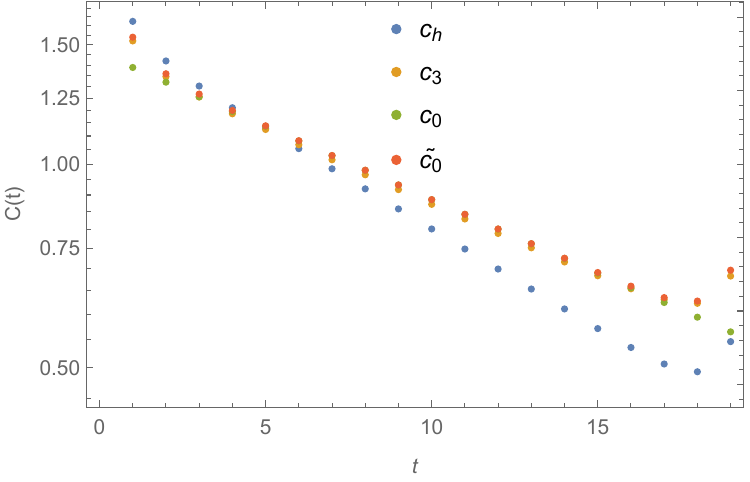}\\
		\includegraphics[width=.45\textwidth]{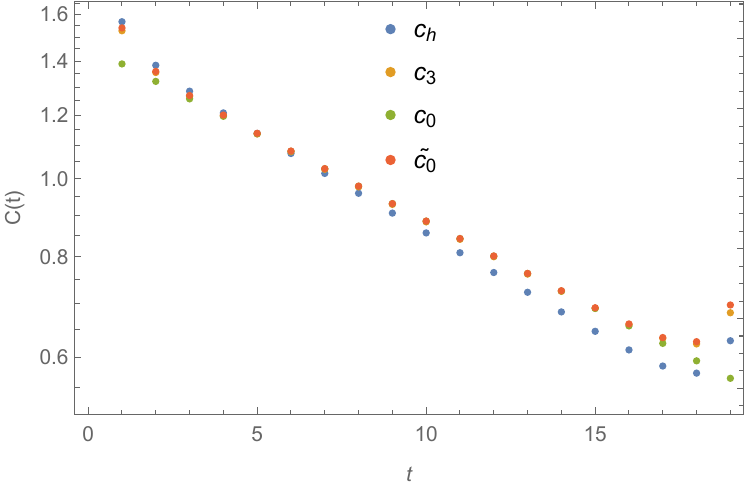}
		\includegraphics[width=.45\textwidth]{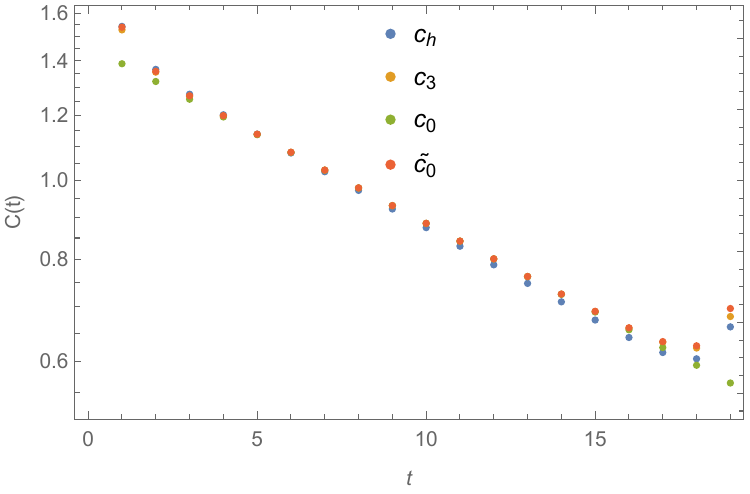}
		\caption{Exact and approximate free fermion propagators at zero spatial momentum in real space, bare mass $m=\num{0.05}$, domain wall separations top: $L_3=3$, $L_s=4$; bottom: $L_s=5$, $L_s=6$.}\label{fig:free_prop_different_Ls}
	\end{figure}

	More importantly, the DW separation $L_s$ has to be chosen finite in actual simulations so that the $L_s\rightarrow\infty$ limit assumed in this section so far is not always justified. We show the behaviour of the propagators at small domain wall separations in figure~\ref{fig:free_prop_different_Ls}. $C_h$ (i.e.\ using $m=m_h$) exhibits significant deviations from the case discussed above for $L\lesssim 6$, whereas $C_3$ (i.e.\ using $m=\im\gamma_3 m_3$) remains virtually unchanged until $L\lesssim 3$.
	We find that both expressions contain first order $\eto{-\alpha L_s}$-terms but $C_3$ has only imaginary and order $\ordnung{m}$ suppressed contributions, resulting in smaller finite $L_s$ effects. See \Cref{sec:dw_height_separation} for the exact formulae.
	
	Let us stress at this point that the finite $L_s$ effects are
significantly larger in the interacting case \added{(Cf.\ the results for the
equation of state presented in \cite{Hands:2020itv,Hands:2021smr})} and one cannot choose the DW
separation from these free theory calculations. We can nevertheless infer that $C_3$ approaches the physical $L_s\rightarrow\infty$ limit faster than $C_h$ justifying our use of this particular formulation throughout this work.
	
\added{
In summary, the results of this section demonstrate that while DWF introduce new
forms of short-distance artifact due to a branch cut not present in traditional lattice
formulations, the results are in perfect accord with the free fermions in the
continuum $am\to0$ and low temperature $L_t\to\infty$ limits. These insights
will aid interpretation of the numerical results for interacting fermions to
follow in Secs.~\ref{sec:results},\ref{sec:conformal}.}

\section{Results from $16^2\times48$}
\label{sec:results}

\begin{figure}[ht]
\begin{center}
\includegraphics[width=0.9\textwidth]{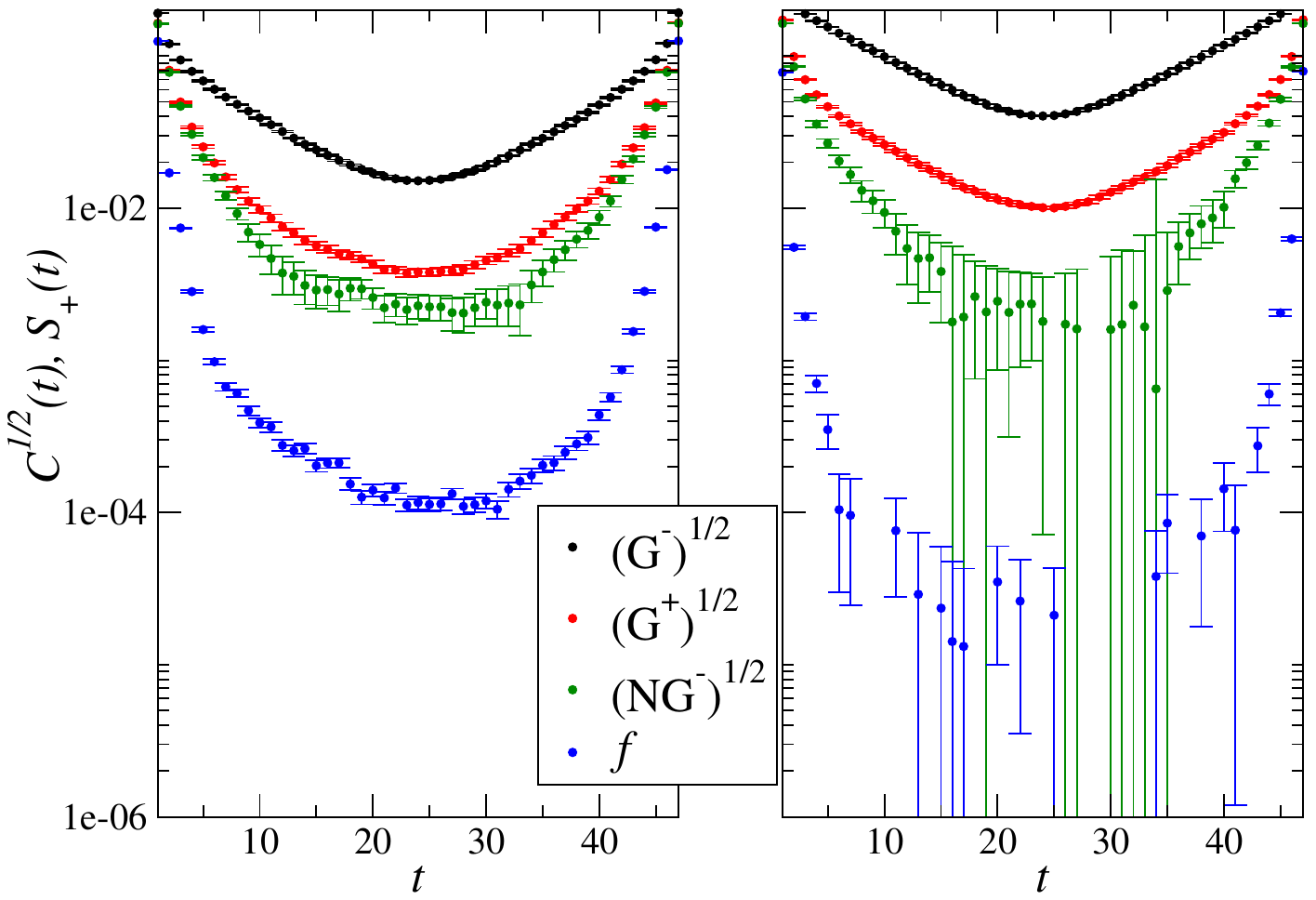}\\
\caption{Timeslice correlators for
$\beta=0.36$ (left) and 0.24 (right) with $ma=0.005$, $L_s=64$.}
\label{fig:correlators}
\end{center}
\end{figure}

In this section we present spectroscopy results from a $16^2\times 48$ system,
using domain wall separations $L_s=64,80$.  In this initial study we have
focussed attention on four values of the inverse coupling $\beta\equiv
g^{-2}a\in\{0.24, 0.28, 0.32, 0.36\}$. For the $N=1$ model defined by
(\ref{eq:lattice_thirring}), the U(2) symmetry
spontaneously breaks at a critical coupling around $\beta\approx0.28$~\cite{Hands:2020itv},\cite{Hands:2021smr},
so we have one ensemble well within the broken phase, two in the symmetric
phase, and one in the vicinity of the critical point. 
Fig.~\ref{fig:correlators} shows data taken from 2500 RHMC trajectories at the
weakest coupling $\beta=0.36$ (left) characterising unbroken symmetry, and the strongest 
$\beta=0.24$ (right) characterising the broken phase.

The data plotted corresponds to the timeslice correlators in G$^-$ (\ref{eq:meson_corr_g5}), G$^+$
(\ref{eq:meson_corr_id}) and NG$^-$ (\ref{eq:meson_corr_g3g5}) meson channels,
and the forwards-moving spin-${1\over2}$ quasiparticle state given by $S_+=S_0+S_3$
(\ref{eq:quasiprop}), denoted $f$ in the figure. Since NG$^+$ also has
contributions from disconnected fermion line diagrams not calculated here, we
merely comment that numerically the connected component is very close to the G$^-$ channel (indeed
they are exactly degenerate in the $m\to0$ limit), and
omit this channel from subsequent analysis. Also note we have chosen to plot the square
root of the meson data in Fig.~\ref{fig:correlators} for ease of comparison with $f$.

As might be anticipated from the form of (\ref{eq:meson_corr_g5}), G$^-$ yields
numerically the largest signal, which increases going from symmetric to broken
phases as first noted in \cite{Hands:2018vrd}. A striking feature is the
disparity between G$^\pm$ channels, which should be degenerate if
U(2) symmetry is manifest. The NG$^-$ data is appreciably noisier, since the
signal (\ref{eq:meson_corr_g3g5}) results from the difference of two much larger
numbers. Finally, $f$ is not symmetric under
$t\mapsto-t$, a generic feature of fermion correlators. For $\beta=0.36$ the $f$
correlator has kink discontinuities about $t=6, L_t-6$ which are compatible
with the branch cut artifacts in the free fermion correlator revealed in the
difference between $C_0$ and $C,\tilde C_0$  in
Fig.~\ref{fig:free_prop_real_space}\added{, discussed in Sec.~\ref{sec:free}}.  At the weaker coupling
$\beta=0.36$ the decay in the forwards $t$-direction  is comparable in all
channels, modulo an overall normalisation, suggesting the mesons are weakly bound states with $M_{\rm
meson}\sim2 M_f$. At $\beta=0.24$ it is possible to discern a difference between
G and NG channels, but by now both NG and $f$ signals are much noisier.
The visible curvature in all data, particularly those at weak coupling, suggests that a
fit assuming conventional exponential decay resulting from an isolated simple
pole may not capture all the information present. Nonetheless, \added{as a first
step} in the next
subsection we will pursue this strategy.

\subsection{Correlator and plateau fits}

We allow two {\em Ans\" atze\/} for the correlator. In the massive case we assume the usual exponential behaviour with (symmetric meson correlator) and without (non-symmetric fermion correlator) back-propagating part
\begin{align}
	C_\text{sym}(t) &= a\cosh(m_\text{eff}(t-L_t/2))\,,\label{eq:correlator_cosh}\\
	C_\text{exp}(t) &= a\,\eto{-m_\text{eff}t}\,,\label{eq:correlator_exp}
\end{align}
respectively, while for fermions with $m=0$ we additionally test for
compatibility with an algebraic decay
\begin{align}
	C_\text{alg}(t) &= \alpha t^{-\mu_\text{eff}}.\label{eq:correlator_algebraic}
\end{align}
In both cases the proportionality constants $a$, $\alpha$ do not carry physical
meaning, whereas the \deleted{`}effective masses\deleted{'} $m_\text{eff}$ and
\deleted{'}anomalous dimension\deleted{'} $\mu_\text{eff}$ are
to be determined, respectively.

To this end we use the procedure derived in Ref.~\cite{Ostmeyer_2020}, Appendix B and summarised in Algorithm 1 thereof.
First, we calculate local approximations of the effective masses
\begin{align}
	m_\text{eff}(t) &= \acosh\left(\frac{C_\text{sym}(t+1)+C_\text{sym}(t-1)}{C_\text{sym}(t)}\right)\,,\label{eq:eff_mass_cosh}\\
	m_\text{eff}(t) &= -\ln\frac{C_\text{exp}(t+1)}{C_\text{exp}(t)}\,,\label{eq:eff_mass_exp}\\
	\mu_\text{eff}(t) &= -\frac{\ln\frac{C_\text{alg}(t+1)}{C_\text{alg}(t)}}{\ln\frac{t+1}{t}}\label{eq:eff_mass_algebraic}
\end{align}
and identify plateaus of the effective mass. Next, we fit a
constant to the plateau in this region and simultaneously one of
formulae~\eqref{eq:correlator_cosh} or~\eqref{eq:correlator_algebraic} directly
to the respective correlator in the same region. The constant plateau fit might
have a bias (see~\cite{Ostmeyer_2020}, or for more details Sec.~4.C
of~\cite{Ostmeyer_2021}), so further analysis always relies on the correlator
fit exclusively. Finally, if the effective mass is not too noisy, we identify
all regions where its slope is compatible with zero, repeat the fit and use the
standard deviation over the different regions' fit results as an estimator of
the systematic error $\Delta_{\rm syst}$.

\begin{figure}[ht]
	\centering
	\includegraphics[width=.45\textwidth,page=97]{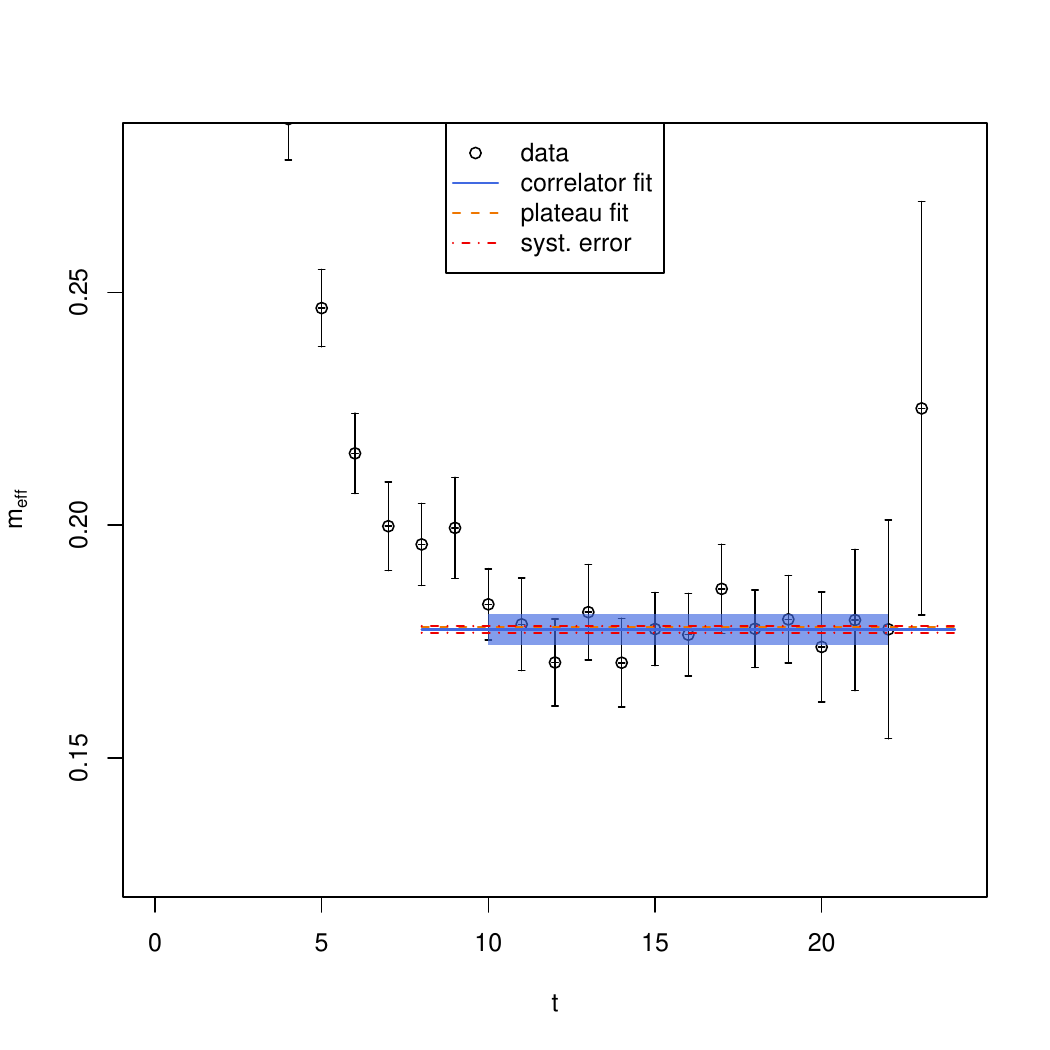}
	\includegraphics[width=.45\textwidth,page=133]{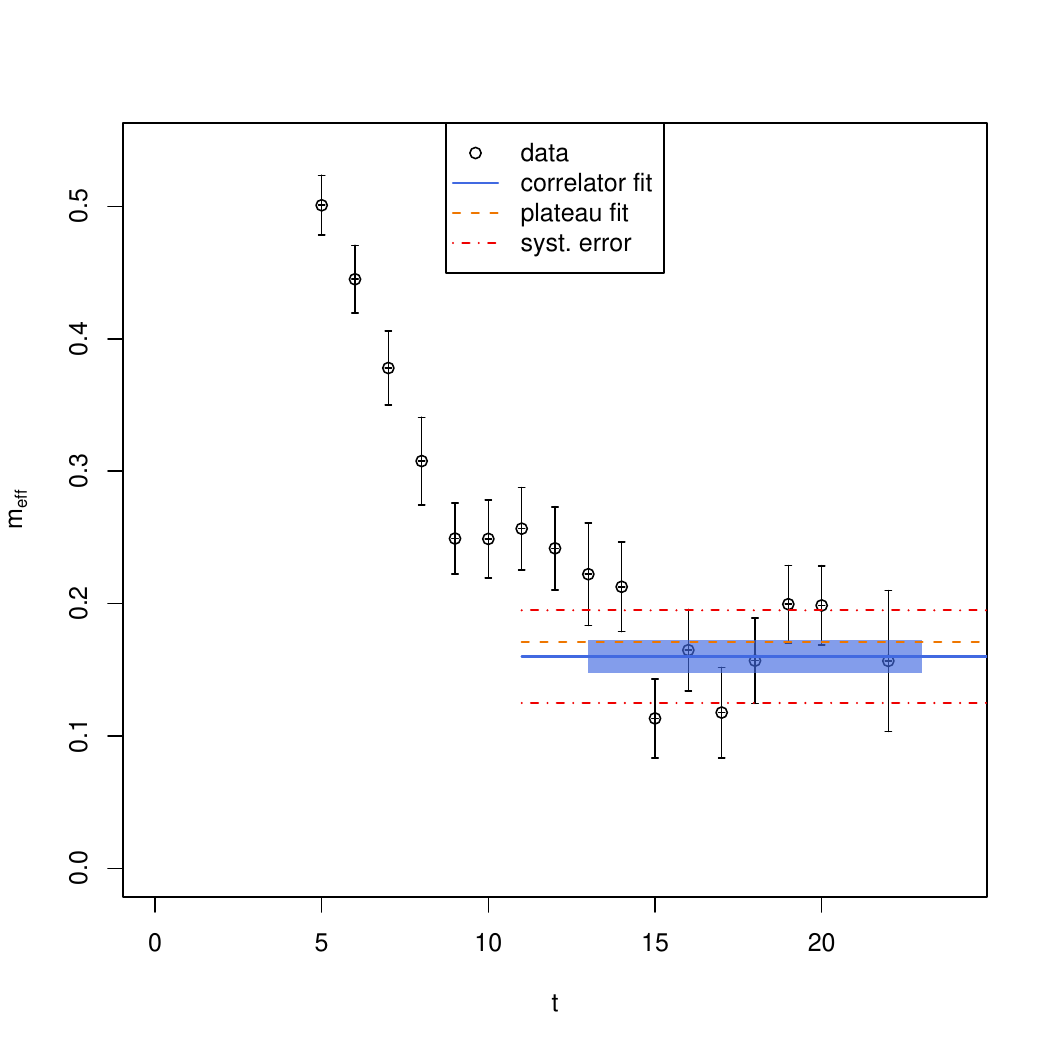}
	\caption{Visualisations of the effective mass~\eqref{eq:eff_mass_cosh} plateaus of $\cosh$-type mesonic `Goldstone' $C_{\One}(x) = C_{\rm G^+}(x)$~\eqref{eq:meson_corr_id} correlator fits.
		The blue line with error band gives result of the correlator fit with statistical error, obtained via eq.~\eqref{eq:correlator_cosh}.
		The length of the blue band indicates the fitting region. For comparison, a constant fit to the effective mass is shown by the dashed orange line. The dot-dashed red line shows the estimation of the systematic error, as explained in~\cite{Ostmeyer_2020}.
		Note that the red and orange lines have been extended outside of the fitting region, for clearer visibility.
		Left: $m=\num{0.005}$, $\beta=\num{0.28}$, $L_s=80$. Right: $m=\num{0.005}$, $\beta=\num{0.36}$, $L_s=64$.}\label{fig:meson_plateau}
\end{figure}

\begin{figure}[ht]
	\centering
	\includegraphics[width=.45\textwidth,page=43]{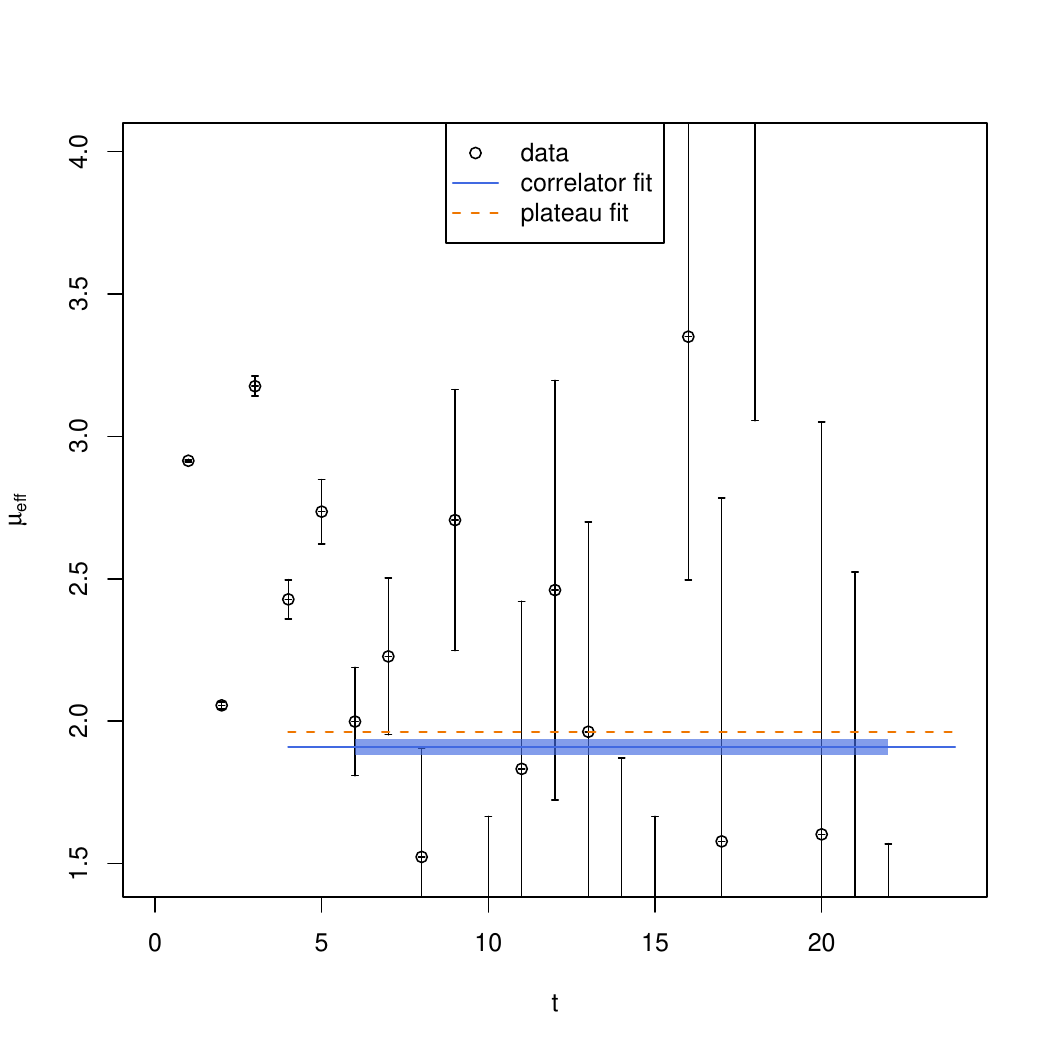}
	\includegraphics[width=.45\textwidth,page=44]{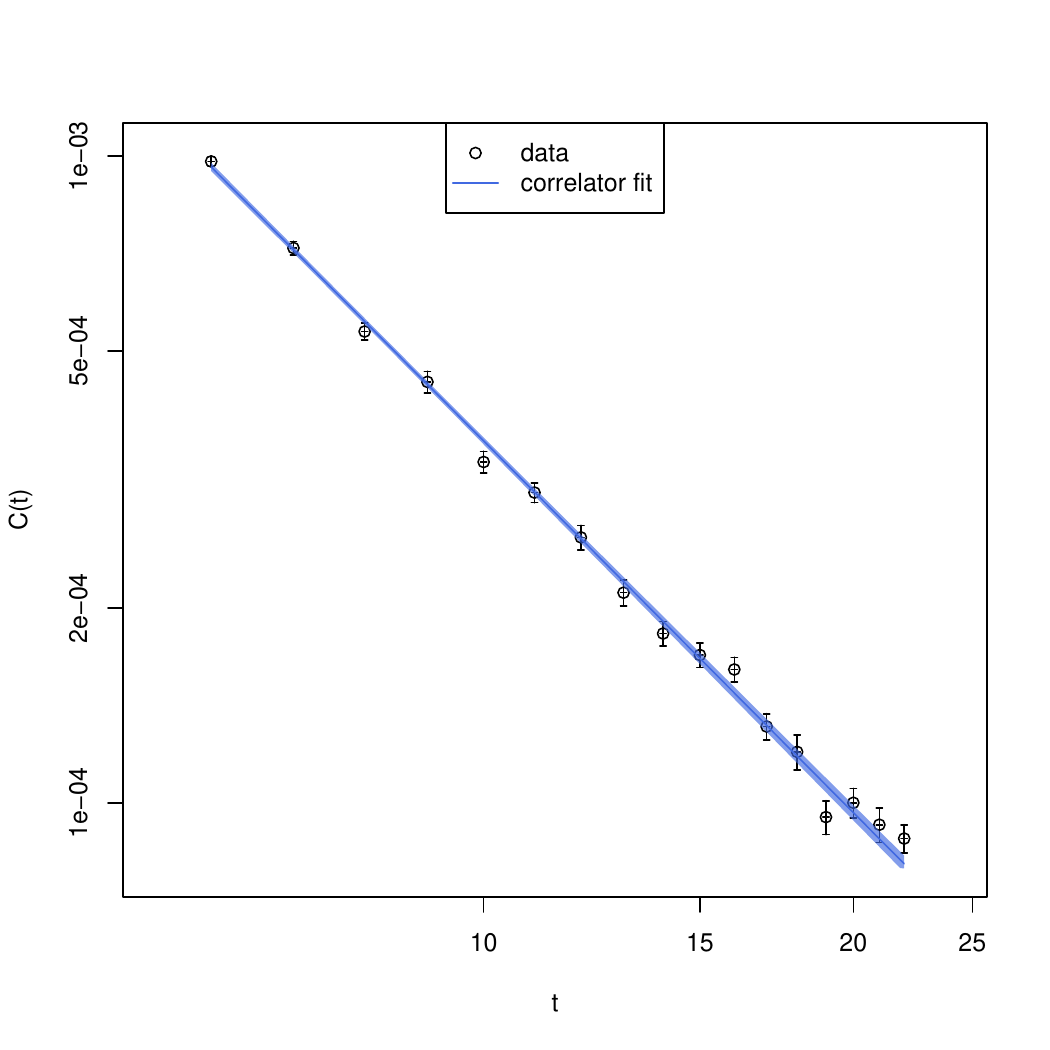}
	\caption{Algebraic fit of a fermionic correlator~\eqref{eq:correlator_algebraic} ($m=\num{0}$, $\beta=\num{0.34}$, $L_s=64$). Visualisation of the effective mass~\eqref{eq:eff_mass_algebraic} (left) and actual fit of the correlator (right). Fit results including statistical errors are shown in blue.
	The length of the blue band indicates the fitting region. For comparison, in the left panel a constant fit to the effective mass is shown by the dashed orange line.}\label{fig:fermion_plateau}
\end{figure}

Figure~\ref{fig:meson_plateau} shows examples of plateaus corresponding to a
weakly and a strongly interacting `Goldstone' meson correlator respectively.
Clearly, the case of $\beta=\num{0.24}$ features a distinct plateau, resulting
in small errors. In contrast, $\beta=\num{0.36}$ comes without an obvious flat
region. This property is captured in a much larger systematic error, 
as seen in Table~\ref{tab:deltasyst} below.

For fermions the effective mass often turns out to be too noisy to be of any
use, as can be seen in the left panel of figure~\ref{fig:fermion_plateau}.
Nevertheless a fit to the correlator is well behaved in most cases (see right
panel of fig.~\ref{fig:fermion_plateau}), so that we can safely analyse the fit
result, albeit without an estimator of potential systematic errors.

\subsection{Results}

\begin{figure}[ht]
\centerline{\includegraphics[width=0.9\columnwidth]{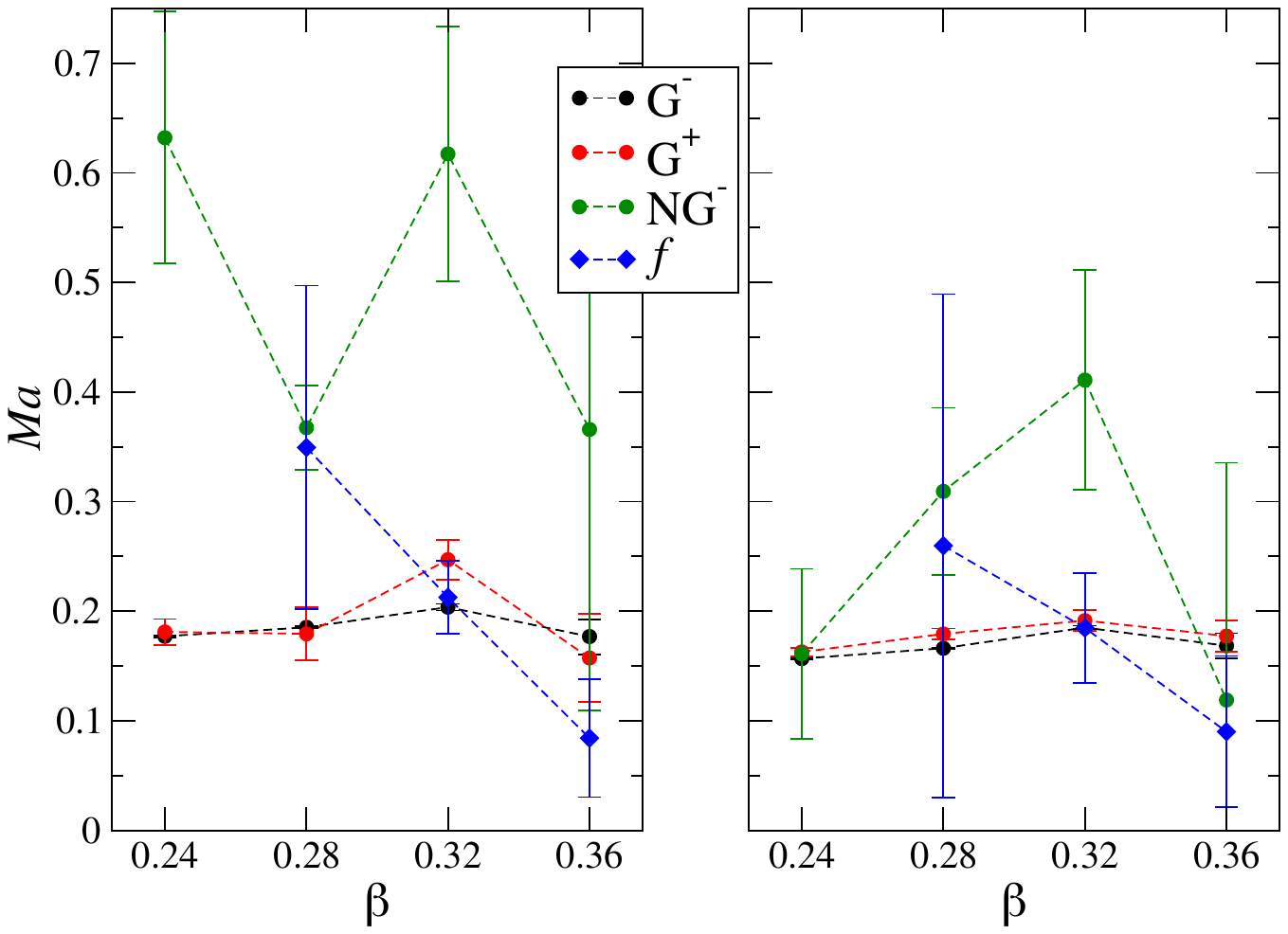}}
\caption{Spectrum results with $ma=0.005$, $L_s=64$ (left) and $L_s=80$ (right). Error
bars are obtained by adding $\Delta_{\rm stat}$ and $\Delta_{\rm syst}$ in
quadrature.}
\label{fig:spectrum}
\end{figure}
Fig.~\ref{fig:spectrum} shows the resulting spectrum in the four channels of
interest for $L_s=64$ (left) and $L_s=80$ (right), using the bare fermion mass
$ma=0.005$, and assuming exponential decay. 
Although there  are relatively large uncertainties in NG$^-$ and $f$ channels,
the picture remains consistent as $L_s$ increases from 64 to 80. The two
G channels yield roughly constant masses across the range of
couplings explored; moreover despite the large disparity in signal amplitude
apparent in Fig.~\ref{fig:correlators}, the G$^\pm$ masses are approximately
degenerate consistent with U(2) symmetry. The $f$ mass satisfies
$M_f\simeq{1\over2}M_{\rm meson}$ at the weakest coupling, but rises sharply
across the critical region $\beta\sim0.28$, consistent with dynamical mass
generation associated with the spontaneous breaking of U(2). No satisfactory
fits were found for the noisy broken phase data at $\beta=0.24$.
The NG$^-$
results are very noisy, but are at least consistent with $M_{\rm NG}\propto M_f$ as
befits a generic non-Goldstone bound state. 

\begin{table}[htp]
\centering
\sisetup{table-format=1.4, table-auto-round}
\begin{tabular}{l|c|c}
$\beta$ & $\Delta_{\rm syst}(L_s=64)$ & $\Delta_{\rm syst}(L_s=80)$\\\hline
0.24 & 0.0008 & 0.0007 \\
0.28 & 0.0001 & 0.0002 \\
0.32 & 0.0029 & 0.0003 \\
0.36 & 0.0158 & 0.0113 \\
\end{tabular}
\caption{Systematic fitting uncertainties in the G$^-$ channel with
$ma=0.005$.}
\label{tab:deltasyst}
\end{table}
As mentioned above, as a consequence of the curvature of the data in the plots
of Fig.~\ref{fig:correlators}, single-pole fits of the form (\ref{eq:correlator_cosh}) are
more convincing in the broken phase, and work less well in the weak-coupling
symmetric phase; this is corroborated by the growth $\Delta_{\rm
syst}$ with $\beta$ exemplified by G$^-$ data shown in
Table~\ref{tab:deltasyst}.
Mesons at weak coupling are weakly-bound
at best, and ultimately may be better described using a continuum spectral
function.

Qualitatively, the picture is very similar to that found in simulations of the
Thirring model with $N=1$ staggered fermions (see Fig.\ 17
of \cite{DelDebbio:1997dv}), in which case the symmetry breaking pattern is U(1)$\otimes$U(1)$\to$U(1). 
For DWF with finite $L_s$ it is necessary to enquire to what extent the
anticipated pattern U(2)$\to$U(1)$\otimes$U(1) is realised.
\begin{figure}[ht]
\centerline{\includegraphics[width=0.9\columnwidth]{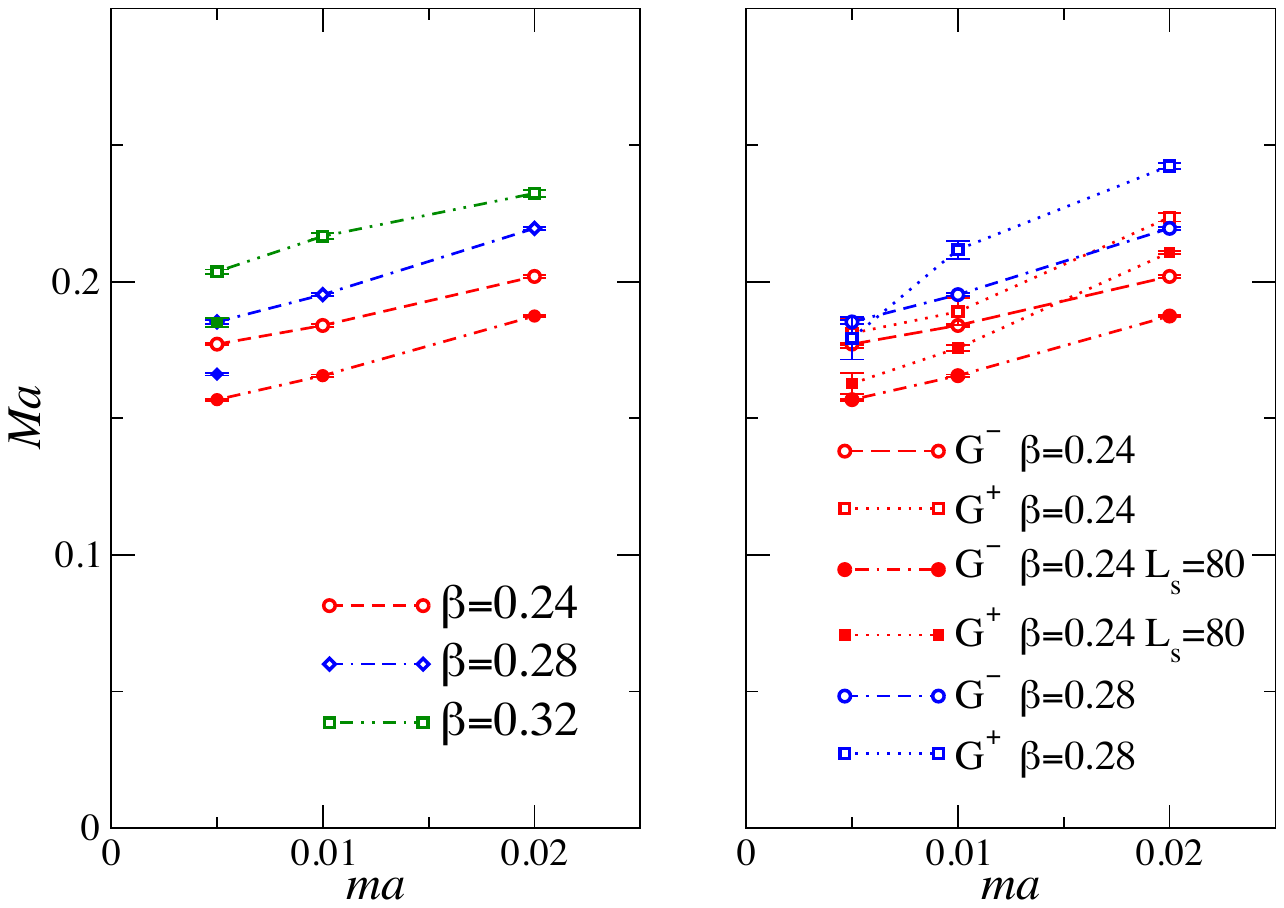}}
\caption{$M_{{\rm G}^-}$ vs. $m$ for various couplings (left); comparison of
$M_{{\rm G}^\pm}$ vs. $m$ (right). In both cases open symbols denote $L_s=64$, closed
$L_s=80$.}
\label{fig:goldstone_scaling}
\end{figure}
Fig.~\ref{fig:goldstone_scaling} addresses this issue from two directions. On
the left is plotted the Goldstone versus bare fermion masses. 
Although $M_{\rm G}^{-}$ decreases with $m$ at all couplings, there is no sign of the $M_{\rm
G}\propto\surd m$ behaviour of a true Goldstone mode in the broken
phase $\beta=0.24$. Comparison of $L_s=64,80$ also suggests the results are not
yet in the large-$L_s$ limit where U(2) recovery is expected. The plot on the
right compares data from the two Goldstone channels G$^\pm$, which with
U(2) symmetry manifest should be
degenerate even for $m\not=0$. At best degeneracy looks to be recovered only as
$m\to0$, and again there are significant finite-$L_s$ effects.
We conclude the results obtained in the meson sector are
suggestive but not yet conclusive, and that U(2) symmetry recovery is not
\added{yet} demonstrated. 

\added{
In summary, in this section we have demonstrated: the presence of meson bound
states (unambiguously in the broken phase $\beta<\beta_c$, more equivocally in
the symmetric phase $\beta>\beta_c$); degeneracy of the two distinct
Goldstone states in pseudoscalar and scalar channels, despite the large
numerical disparity in the correlators; the expected hierarchy between G and
NG states; the evolution in fermion mass from weak coupling where
mesons are weakly-bound $f\bar f$ states to strong coupling
where there is dynamical gap generation and $M_f$ is hard to measure.
The scaling of the Goldstone masses with bare fermion mass does not manifest the
expected $M_G\propto\surd m$ behaviour, and further work to explore both
thermodynamic and large-$L_s$ limits is needed.}

\section{Conformal Nature of the Fermion Correlator}
\label{sec:conformal}

\newcommand{\lapprox}{\raisebox{-0.5ex}{$\
\stackrel{\textstyle<}{\textstyle\sim}\ $}}
\newcommand{\gapprox}{\raisebox{-0.5ex}{$\
\stackrel{\textstyle>}{\textstyle\sim}\ $}}

While spectroscopy with explicit U(2) symmetry-breaking $m\not=0$ is the best
way to test the Goldstone nature of the bound states G$^\pm$, it does not reveal
the critical nature of the fermion at the fixed point. In this section we
discuss the case $m=0$, presenting both continuum-based models for the critical propagator,
characterised by a new and distinct exponent $\eta_\psi$, and numerical data for
the propagator $S_0$ (\ref{eq:quasiprop}) taken in the massless limit.

\subsection{Massless fermions}
\added{To begin,} \deleted{In this section} we propose a model for the fermion timeslice correlator
$C_f(x_0)$ in the symmetric
phase $g^2<g_c^2$ in the massless limit $m\to0$. In this regime we expect the
correlator to decay algebraically, but also to reflect in some way a finite
correlation length which diverges only as $g^2\to g^2_{c-}$. 
Our ultimate aim is to identify the fermion anomalous
dimension $\eta_\psi$ defined by the critical scaling
\begin{equation}
C_f(\vec p)\sim{{\hat{p}\cdot\vec\gamma}\over{\vert\vec p\vert^{1-{\eta_\psi}}}}
\;\;\;\Leftrightarrow\;\;\;
C_f(\vec x)\sim{{\hat{x}\cdot\vec\gamma}\over{\vert\vec x\vert^{2+{\eta_\psi}}}}
\label{eq:conformal}
\end{equation}
with $\vec\gamma=(\gamma_0,\gamma_1,\gamma_2)$, $\hat x\cdot\vec x=\vert\vec
x\vert\equiv x$.

We start by focussing on the behaviour exactly at the critical point, modelled by
replacing the free massless fermion momentum-space propagator $1/ip{\!\!\! /}$
by $1/ip^{1-\eta_\psi}\hat p_\mu\gamma_\mu$ with $\vert\hat p\vert=1$ and 
$\hat
p_\mu p_\mu=p$: 
\begin{eqnarray}
C_f(x)&=&{\rm tr}\left\{{\gamma_{\hat x}\over4}\int
{{d^3p}\over{(2\pi)^3}}{{-i\hat p_\mu\gamma_\mu}\over
{p^{1-\eta_\psi}}}e^{i\vec p\cdot\vec x}\right\}\nonumber\\
&=&\int {dp\over8\pi^2}
p^{1+\eta_\psi}\int_{-{\pi\over2}}^{\pi\over2}d\theta\sin2\theta\sin(px\cos\theta)\nonumber\\
&=&{x^{-{1\over2}}\over{(2\pi)^{3\over2}}}\int_0^\infty
dpp^{{1\over2}+\eta_\psi}J_{3\over2}(px).\label{eq:integral}
\end{eqnarray}
The remaining integral over $p$ is formally given by
\begin{equation}
C_f(x)={1\over{4\pi
x^{2+\eta_\psi}}}{{\Gamma(2+\eta_\psi)}\over{\Gamma(1+{{\eta_\psi}\over2})\Gamma(1-{{\eta_\psi}\over2})}};\;\;\;
\lim_{\eta_\psi\to0}C_f(x)={1\over{4\pi x^2}}.
\label{eq:algebraic}
\end{equation}
Since the decay is algebraic, it is natural to plot $C_f(x)$ using
logarithmic scales on both $x$ and $y$-axes.

\subsection{UV considerations}

The integral (\ref{eq:integral}) is only convergent for $\eta_\psi<0$: in general therefore
we must introduce a UV scale $\Lambda$ to regularise the model. 
A simple sharp momentum-space cutoff $p\leq\Lambda$ yields an oscillatory
dependence $C_f(x)\propto\cos(\Lambda x)$, which is physically unacceptable. We have
explored a smoother cutoff defined by the following integral, which
exists for $\eta_\psi>-3$:
\begin{equation}
{x^{-{1\over2}}\over{(2\pi)^{3\over2}}}\int_0^\infty
dpp^{{1\over2}+\eta_\psi}J_{3\over2}(px)e^{-{p^2\over\Lambda^2}}=
{{\Gamma({{3+\eta_\psi}\over2})}\over{12\pi^2}}x\Lambda^{3+\eta_\psi}
M\left(\textstyle{{{3+\eta_\psi}\over2};{5\over2};-{{x^2\Lambda^2}\over4}}\right),
\label{eq:Lambda2}
\end{equation}
where $M$ is the confluent hypergeometric
function  $_1F_1$. In the limit $x\Lambda\to\infty$
(\ref{eq:Lambda2}) recovers the naive algebraic decay 
(\ref{eq:algebraic}).

\begin{figure}[h]
\begin{center}
  \includegraphics[width=0.9\columnwidth]{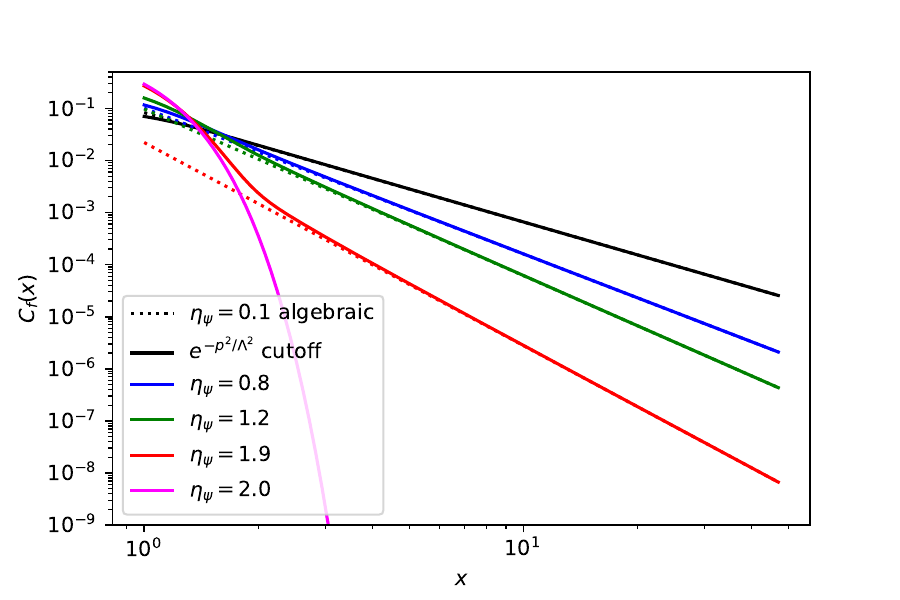}\\
  \caption{ $C_f(x)$ for various $\eta_\psi$ with $\Lambda=\pi$.}
  \label{fig:correlator_UV}
\end{center}
\end{figure}
Fig.~\ref{fig:correlator_UV} shows $C_f(x)$ evaluated using both
(\ref{eq:algebraic},\ref{eq:Lambda2}). To approximate
the lattice cutoff we choose a numerical value $\Lambda=\pi$. For
$\eta_\psi\lapprox1$ the regularised form matches the algebraic form well, but
for larger values of $\eta_\psi$ the cutoff dependence is significant over much
of the
range permitted by $L_t=48$.
As dictated by the gamma function in the denominator of
(\ref{eq:algebraic}), things break down at $\eta_\psi=2$ where
(\ref{eq:Lambda2}) has the limiting form 
\begin{equation}
C_f(x)={{x\Lambda^5}\over{16\pi\surd\pi}}\exp\left(-{{x^2\Lambda^2}\over4}\right),
\end{equation}
and it is no longer possible to hide the cutoff.

We conclude: (i) for conformal dynamics described by (\ref{eq:conformal})
there appears to be an upper bound on the anomalous dimension
$\eta_\psi<2$; (ii) for large anomalous dimensions UV artifacts might make
fitting for $\eta_\psi$ a non-trivial challenge. 

\subsection{Introduction of finite correlation length}
\label{sec:model}
Next we introduce a finite correlation length $\mu^{-1}$, motivated by the
large-$N$ limit
of the scalar auxiliary field propagator found in the 2+1$d$ Gross-Neveu
model~\cite{Kikukawa:1989fw}:
\begin{equation}
D_\sigma(p)={4\mu\over{p+\mu}},\label{eq:GNsigma}
\end{equation}
where the inverse correlation length $\mu$ is related to the width of an unstable resonance
in the scalar channel, but does not correspond to a pole on the imaginary-$p$
axis yielding exponential decay. 
Rather, the appearance of $p=(\vec p\cdot\vec p)^{1\over2}$ in the denominator yields a branch cut starting
at the origin in the complex $p^2$ plane; the pole of (\ref{eq:GNsigma}) at
$(p^2)^{1\over2}=-\mu$ lies
on a different sheet to the one where the integral defining the Fourier
transform from $p$ to $x$ lives. This form was used to fit numerical $D_\sigma$ data in
\cite{Hands:1992be}.

Our {\em Ansatz} for the fermion propagator in momentum
space is
\begin{equation}
C_f(p)={1\over{i(p+\mu)^{1-\eta_\psi}\hat p_\mu\gamma_\mu}}={{-i\hat
p_\mu\gamma_\mu}\over{(p+\mu)^{1-\eta_\psi}}}.
\label{eq:Ansatz}
\end{equation}
In real space we now have
\begin{equation}
C_f(x)={1\over4}{\rm tr}\gamma_{\hat
x}C_f(p)={x^{-{1\over2}}\over{(2\pi)^{3\over2}}}\int_0^\infty
dp(p+\mu)^{\eta_\psi-1}p^{3\over2}J_{3\over2}(px).
\end{equation}
In the limit $\mu\to0$ we recover (\ref{eq:algebraic}), while for
$\mu\to\infty$
\begin{equation}
C_f(x)={1\over{\pi^2\mu^{1-\eta_\psi}x^3}}.
\label{eq:xcubed}
\end{equation}
\begin{figure}[h]
\begin{center}
  \includegraphics[width=0.9\columnwidth]{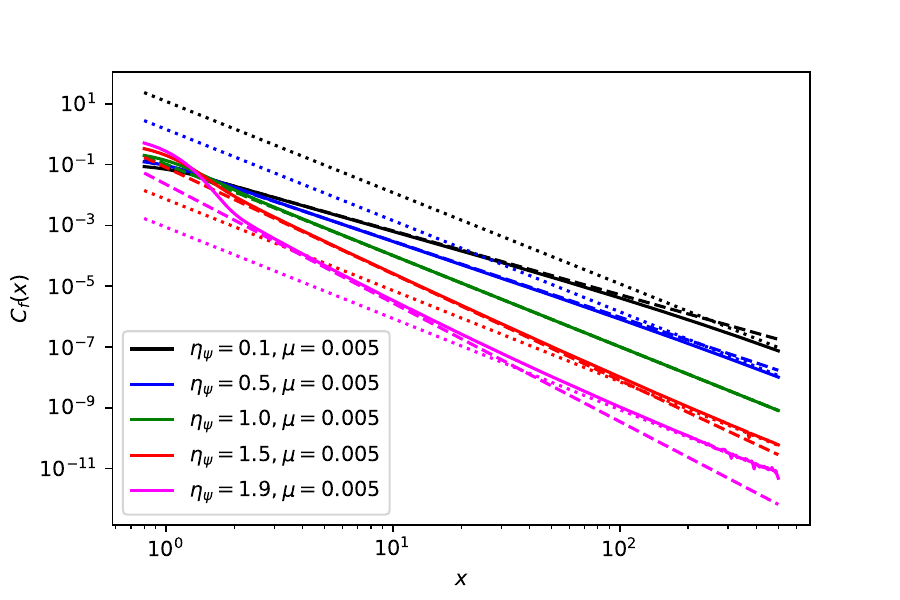}\\
  \caption{ $C_f(x)$ for various $\eta_\psi$ with $\Lambda=\pi$ and $\mu=0.005$.}
  \label{fig:correlator_fixedmu}
\end{center}
\end{figure}
Once again a UV regulator $e^{-{p^2\over\Lambda^2}}$ must be introduced,
which modifies $C_f(x)$ at small $x$.
The resulting integral may be evaluated using numerical quadrature; the result
for fixed $\mu$ and varying $\eta_\psi\in(0,2)$ is shown in
Fig.~\ref{fig:correlator_fixedmu}.
Dashed and dotted lines show the limiting forms (\ref{eq:algebraic}) and
(\ref{eq:xcubed}) respectively.
\begin{figure}[p]
\begin{center}
  \includegraphics[width=0.9\columnwidth]{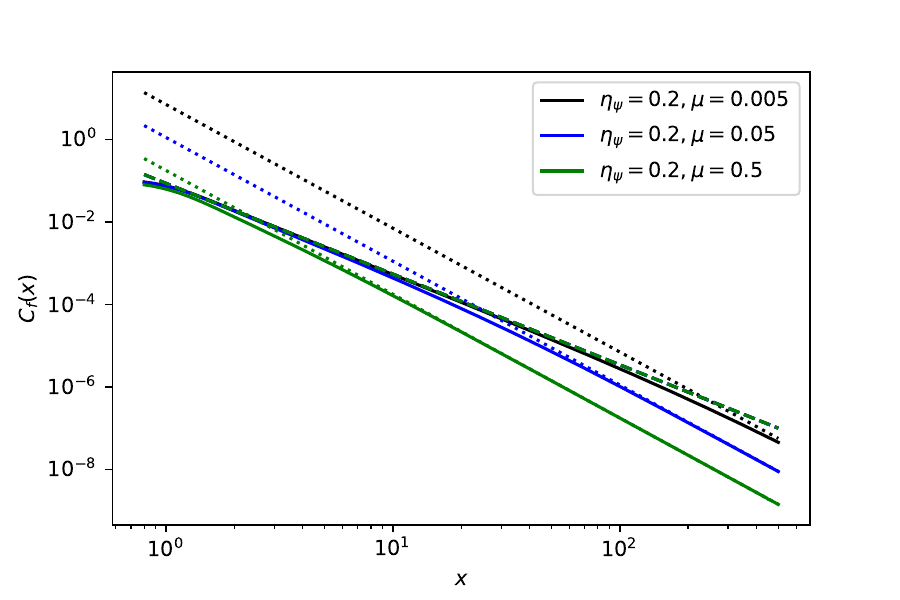}\\
  \caption{ $C_f(x)$ for $\eta_\psi=0.2$ with $\Lambda=\pi$ and various $\mu$.}
  \label{fig:correlator_fixedeta02}
\end{center}
\vskip 1 truecm
\begin{center}
  \includegraphics[width=0.9\columnwidth]{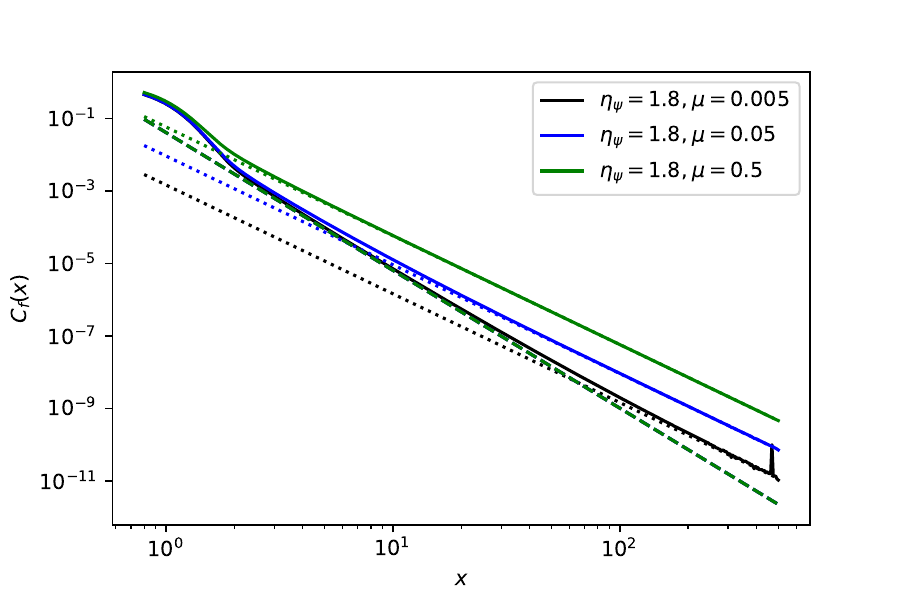}\\
  \caption{ $C_f(x)$ for $\eta_\psi=1.8$ with $\Lambda=\pi$ and various $\mu$.}
  \label{fig:correlator_fixedeta18}
\end{center}
\end{figure}
 
In practice, our dataset is hypothesised to have fixed $\eta_\psi$ and varying
$\mu$ corresponding to varying $g_c^{-2}-g^{-2}$. 
Figs.~\ref{fig:correlator_fixedeta02},~\ref{fig:correlator_fixedeta18},
show $C_f(x)$ evaluated for various $\mu$ with fixed $\eta_\psi$ chosen close
to the extremes of the range (0,2). The curvature of the plots suggests it may be
possible to distinguish the cases $\eta_\psi<1$ and $\eta_\psi>1$ by qualitative
means without recourse to a fitting analysis where control of systematics is
still poorly understood. 

However, the Monte Carlo data is for the
timeslice correlator $C_{ft}(x_0)$. For $C_f(x)=\hat
x\cdot \vec\gamma/x^{2+\eta_\psi}$, 
\begin{equation}
C_{ft}(x_0) ={1\over4}{\rm tr}\gamma_0\int d^2x_\perp C_f(x)
={{2\pi}\over{1+\eta_\psi}}{1\over
x_0^{\eta_\psi}}.\label{eq:finiteV}
\end{equation}
We therefore predict the slope of the resulting data on a log-log plot to be
in the range (0,2) for $\Lambda x_0\gg1$, $\mu x_0\ll1$, with asymptotic slope 1 achieved for $\mu x_0\gg1$. 

\subsection{Numerical results}

We calculated the fermion timeslice propagator with $m=0$ using just the
time-symmetric projection
$S_0$ of (\ref{eq:quasiprop}) on ensembles with $L_s=64$ generated by 5000 RHMC trajectories
taken at 5 $\beta$-values in the symmetric phase, with the strongest
$\beta=0.28$ corresponding approximately to the critical  value obtained in
studies of the equation of state~\cite{Hands:2020itv,Hands:2021smr}.
\begin{figure}[h]
\begin{center}
\includegraphics[width=12.0cm]{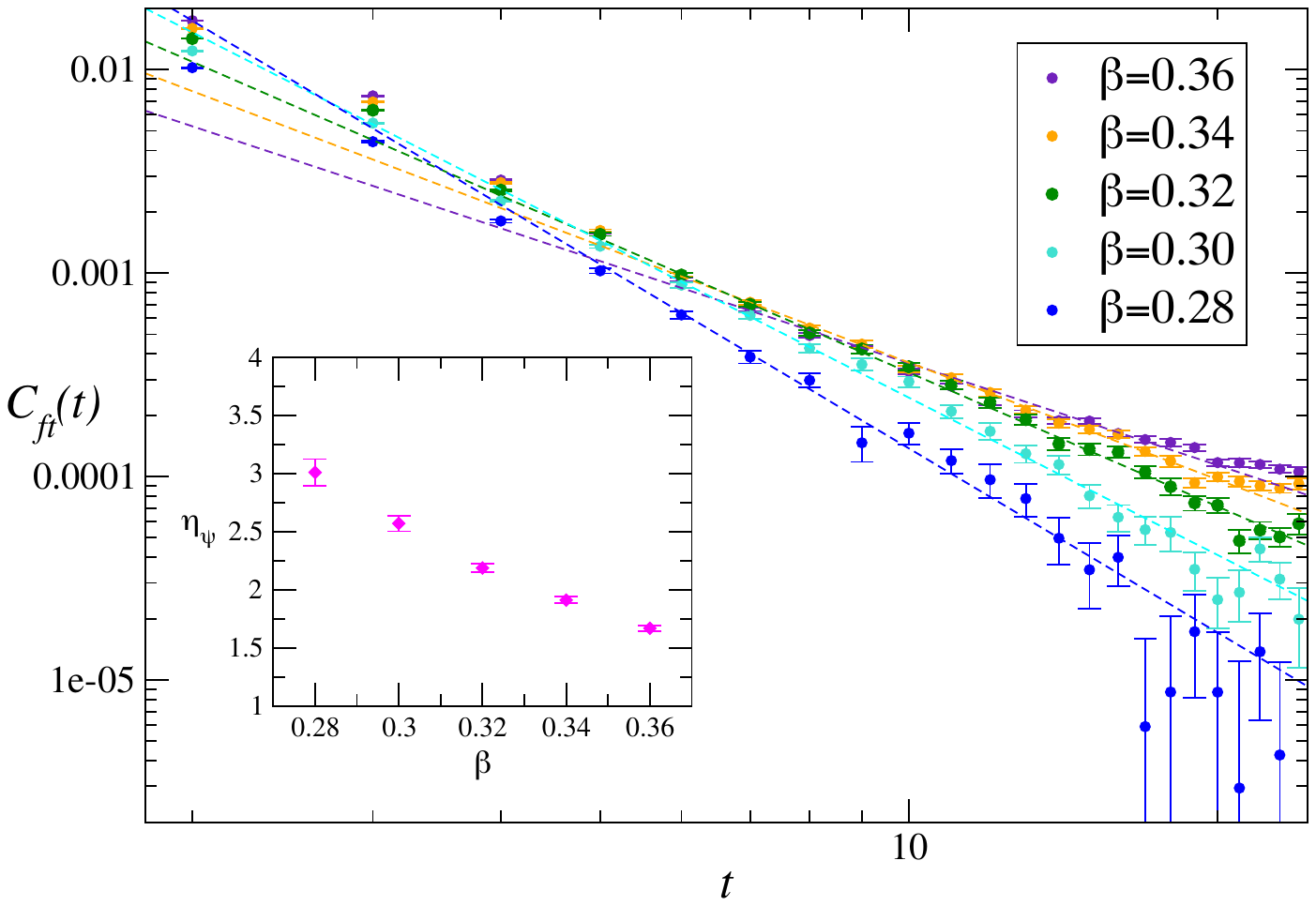}
\caption{Timeslice fermion correlator for various $\beta$ with $L_s=64$, $m=0$,
together with fits to (\ref{eq:correlator_algebraic}). The inset shows the fitted anomalous dimension
$\eta_\psi$.} \label{fig:ferm_alg}
\end{center}
\end{figure}
The results for data averaged over forwards and backwards directions are shown on a log-log plot in Fig.~\ref{fig:ferm_alg}.
Whilst the data show qualitative features which might be compared with those
of Fig.~\ref{fig:correlator_fixedeta18} describing the case $\eta_\psi>1$, the signal-to-noise ratio falls as the
critical coupling is approached.

In principle on a lattice of finite temporal extent we should correct for 
backwards propagating signals, and also contributions
from propagation over arbitrarily many temporal circuits.
The analysis presented in \Cref{sec:IR} shows that both effects are
mitigated by fermion antiperiodic boundary conditions; \added{however}, with the
limited statistical precision
currently achieved there is no motivation to explore  beyond the
simplest fit form (\ref{eq:correlator_algebraic}). Results for the fitted
$\eta_\psi=\mu_{\rm eff}$ are 
plotted in the inset of Fig.~\ref{fig:ferm_alg}.
The quality of fit increases as $\beta\searrow\beta_c$ and the fitted
$\eta_\psi\simeq3$ for the near-conformal value $\beta=0.28$.

\added{In summary, we have established a signal and 
obtained results for the fermion timeslice correlator in the
vicinity of the postulated QCP in the massless limit, and analysed its 
propagation assuming algebraic decay in the temporal direction. Our results are
in qualitative agreement}
\deleted{Our results are inconsistent} with the model presented in
Sec.~\ref{sec:model}\added{, but disagree quantitively} in two respects:
firstly the value for the fermion anomalous dimension
$\eta_\psi\deleted{\approx3}$ lies
outside the range [0,2) compatible with the 
conformal propagator in momentum space
(\ref{eq:conformal}); secondly there is no sign of recovery of $\mu_{\rm
eff}=1$ far from criticality as predicted by
(\ref{eq:Ansatz},\ref{eq:xcubed}). \added{Rather, our data suggest
$\eta_\psi\approx3$ in the vicinity of the QCP, an unexpectedly large value.} 
 
\section{Discussion}
\label{sec:disco}

In this paper we have used the DWF formulation of the Thirring model 
to perform spectroscopy using orthodox lattice field theory techniques 
on a lattice in which the temporal
extension is greater than the spatial extent. Since the Thirring
model has no gauge symmetry, we have also been able to study propagation of
elementary fermion fields. In both cases the results represent significant
progress over previous exploratory studies~\cite{Hands:2016foa,Hands:2018vrd}.

In the meson sector, our results are consistent with the two Goldstone-like
modes G$^\pm$ having  degenerate masses, as demanded by the residual U(1)$\otimes$U(1)
symmetry, despite a large disparity in the overall magnitude of the correlators.
Moreover the accessible non-Goldstone NG$^-$ state is clearly more massive,
and increasing as the system moves from the symmetric into the broken phase,
despite a very noisy signal due to its definition in terms of the difference of
two much larger correlators. At the weakest couplings $\beta=0.36$ explored,
the poor quality of fit reflected in the large systematic error $\Delta_{\rm
syst}$ suggests mesons are only very weakly bound so there is significant contamination
from a fermion-antifermion continuum. In the broken phase at $\beta=0.24$ by
contrast, \deleted{while} the Goldstones are tightly bound\added{, but}
\deleted{they} fail to respect the
expected behaviour $M_{\rm G}\propto\surd m$; the large $L_s$-artifacts shown in
Fig.~\ref{fig:goldstone_scaling} suggest U(2) symmetry
restoration may be even harder to observe in the spectrum than in the order
parameter\added{: studies in the thermodynamic limit, not
taken here, may also prove important}.

In the fermion sector, we have first presented an analysis of the free-field
correlator which highlights a potentially significant contamination arising from
a fermion-doubler continuum at small $t$, \added{clearly visible in the data of
Fig.~\ref{fig:correlators}, which must be taken into account when performing
spectroscopic fits. We} \deleted{and} also examined the impact of
varying domain wall height $M$ and separation $L_s$. Once again, the superior
convergence of the DWF formulation with mass term $S_{m_3}$ (\ref{eq:mass_3}) is apparent. 
A major innovation has been the employment of wall sources to vastly improve the
sampling of the correlator over previous attempts~\cite{Hands:2016foa}. The
results indicate $M_f\approx{1\over2}M_{\rm G} \simeq {1\over2}M_{\rm NG}$
at weak coupling $\beta=0.36$,
consistent with mesons being weakly bound states, but
that $M_f$ rises steeply towards the phase transition at $\beta_c\approx0.28$
until correlator noise precludes its measurement in the broken phase. It is
clear a much finer comb of coupling strengths in the vicinity of $\beta_c$ is
needed in order to refine this rather crude first step.

Finally we exploited the stability of the DWF formulation to perform
measurements of the fermion propagator in the $m\to0$ limit in an attempt to probe
the conformal nature of the putative fixed point dynamics. We first presented an
analytic model demonstrating that in general a UV regularisation is needed, and
that the algebraic decay of the fermion correlator parametrised by the critical
exponent $\eta_\psi$ needs to satisfy the bounds $0\leq\eta_\psi<2$ in order to permit a
straightforward passage between real space and momentum space.
The numerical data of Fig.~\ref{fig:ferm_alg} by contrast prefer
$\eta_\psi\approx3$. We are at a loss to account for this discrepancy, but know
of no reason why a conformal field theory with such a large anomalous
dimension should be excluded. This is certainly the most interesting result of
this paper, and contrasts markedly with the value $\eta_\psi=0.37(1)$ found in
finite volume scaling studies of both Thirring~\cite{Chandrasekharan:2011mn} and
U(1) Gross-Neveu~\cite{Chandrasekharan:2013aya} models
formulated with staggered fermions, strengthening our conviction that DWF and
staggered Thirring models are distinct. 

In conclusion, while \deleted{progress has been made} \added{much insight
into the excitations of the Thirring model as the U(2) limit is approached as $L_s\to\infty$ has been
obtained}, 
\replaced{some important}{many} questions remain
unanswered.  In future work, encouraged by the apparent superior $L_s$
convergence observed in \cite{Worthy:2021ddb}, we plan to extend the study to the Thirring model
formulated with DWF using a Wilson kernel rather than the simplest Shamir kernel
(\ref{eq:Shamir}) used here.
 
\section*{Acknowledgements}

This work was performed using the Cambridge Service for Data Driven Discovery
(CSD3), part of which is operated by the University of Cambridge Research
Computing on behalf of the STFC DiRAC HPC Facility (www.dirac.ac.uk). The DiRAC
component of CSD3 was funded by BEIS capital funding via STFC capital grants
ST/P002307/1 and ST/R002452/1 and STFC operations grant ST/R00689X/1. DiRAC is
part of the National e-Infrastructure. Additional work used the {\em Sunbird\/}
facility of Supercomputing Wales.
SJH was supported in part by the STFC Consolidated Grant ST/T000813/1, and JO by ST/T000988/1.

\section*{Data Access}
\added{
All the code and data (raw as well as analysed) required to reproduce the results of this paper can be found in~\cite{liverpool_rdm1959} under open access. We used \texttt{Fortran} for the DW simulations and heavily relied on the \texttt{hadron}~\cite{hadron} package for our data analysis implemented in \texttt{R}~\cite{r_language}.
}

\appendix
\section{Derivation of the free fermion propagator in time}
\label{sec:mats}
\subsection{List of auxiliary variables}\label{sec:aux_vars}
The auxiliary variables required in the equations~\eqref{eq:corr_same_wall} and~\eqref{eq:corr_diff_walls} are given by
\begin{align}
	2\cosh\alpha &= \frac{1+b^2+\bar p^2}{b}\,; & \bar p_\mu &=\sin p_\mu\,; & b(p) &= 1-M+\sum_\mu (1-\cos p_\mu)
	\label{eq:alpha}
\end{align}
as well as
\begin{align}
	A_\pm &= \Delta^{-1}B\left(\eto{\pm\alpha}-b\right)\left(1-|m|^2\right)\,,\\
	A_m &= \Delta^{-1} B\left[-2mb\sinh \alpha + \eto{-\alpha(L_s-1)}\left(\eto{-2\alpha}(b-\eto{\alpha}) + |m|^2(\eto{-\alpha}-b)\right)\right]\,,\\
	B &= (2b\sinh\alpha)^{-1}\,,\\
	\begin{split}
		\Delta &= \eto{2\alpha}(b-\eto{-\alpha})+|m|^2(\eto{\alpha}-b)\\
		&\quad + \eto{-2\alpha(L_s-1)}\left[|m|^2(b-\eto{-\alpha})+\eto{-2\alpha}(\eto{\alpha}-b)\right]\\
		&\quad + \delta_{m,m_h}\,\eto{-\alpha(L_s-1)}4mb\sinh\alpha\,.
	\end{split}
\end{align}
The mass can be chosen to be either $m=m_h$ or $m=\im m_3$.

\subsection{Holomorphic case up to poles}\label{sec:c0_derivation}
\begin{figure}[ht]
		\centering
\begingroup%
  \makeatletter%
  \providecommand\color[2][]{%
    \errmessage{(Inkscape) Color is used for the text in Inkscape, but the package 'color.sty' is not loaded}%
    \renewcommand\color[2][]{}%
  }%
  \providecommand\transparent[1]{%
    \errmessage{(Inkscape) Transparency is used (non-zero) for the text in Inkscape, but the package 'transparent.sty' is not loaded}%
    \renewcommand\transparent[1]{}%
  }%
  \providecommand\rotatebox[2]{#2}%
  \newcommand*\fsize{\dimexpr\f@size pt\relax}%
  \newcommand*\lineheight[1]{\fontsize{\fsize}{#1\fsize}\selectfont}%
  \ifx\svgwidth\undefined%
    \setlength{\unitlength}{381.00782545bp}%
    \ifx\svgscale\undefined%
      \relax%
    \else%
      \setlength{\unitlength}{\unitlength * \real{\svgscale}}%
    \fi%
  \else%
    \setlength{\unitlength}{\svgwidth}%
  \fi%
  \global\let\svgwidth\undefined%
  \global\let\svgscale\undefined%
  \makeatother%
  \begin{picture}(1,0.72837901)%
    \lineheight{1}%
    \setlength\tabcolsep{0pt}%
    \put(0,0){\includegraphics[width=\unitlength,page=1]{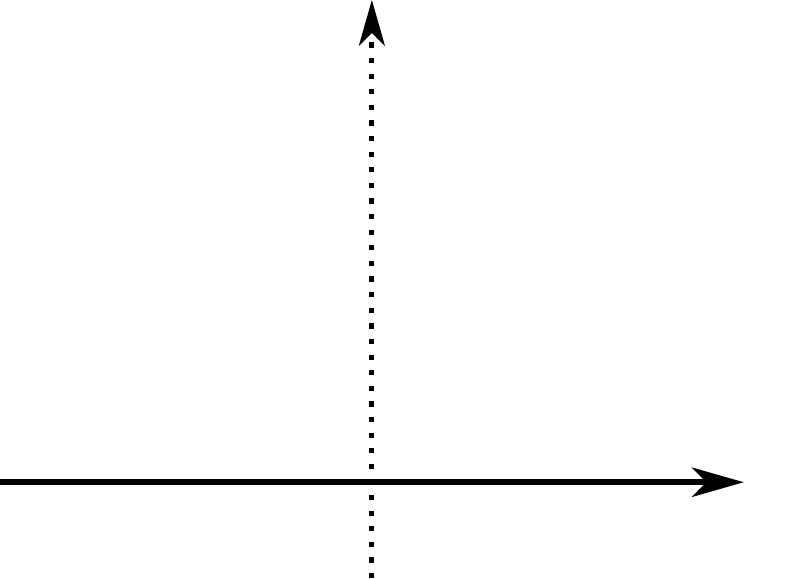}}%
    \put(0.87583045,0.15259323){\color[rgb]{0,0,0}\makebox(0,0)[lt]{\lineheight{1.25}\smash{\begin{tabular}[t]{l}$\real(z)$\end{tabular}}}}%
    \put(0.48418607,0.70104043){\color[rgb]{0,0,0}\makebox(0,0)[lt]{\lineheight{1.25}\smash{\begin{tabular}[t]{l}$\imag(z)$\end{tabular}}}}%
    \put(0,0){\includegraphics[width=\unitlength,page=2]{poles+first_contour.pdf}}%
    \put(0.11867174,0.07496385){\color[rgb]{0,0,0}\makebox(0,0)[lt]{\lineheight{1.25}\smash{\begin{tabular}[t]{l}$-\asinh m$\end{tabular}}}}%
    \put(0,0){\includegraphics[width=\unitlength,page=3]{poles+first_contour.pdf}}%
    \put(0.49682615,0.08620447){\color[rgb]{0,0,0}\makebox(0,0)[lt]{\lineheight{1.25}\smash{\begin{tabular}[t]{l}0\end{tabular}}}}%
    \put(0,0){\includegraphics[width=\unitlength,page=4]{poles+first_contour.pdf}}%
    \put(0.50654783,0.55795585){\color[rgb]{0,0,0}\makebox(0,0)[lt]{\lineheight{1.25}\smash{\begin{tabular}[t]{l}$\im 2\pi$\end{tabular}}}}%
    \put(0.49131318,0.21972823){\color[rgb]{0,0,0}\makebox(0,0)[lt]{\lineheight{1.25}\smash{\begin{tabular}[t]{l}$\im\pi/L_t$\end{tabular}}}}%
    \put(0.49131318,0.36852547){\color[rgb]{0,0,0}\makebox(0,0)[lt]{\lineheight{1.25}\smash{\begin{tabular}[t]{l}$3\im\pi/L_t$\end{tabular}}}}%
    \put(0,0){\includegraphics[width=\unitlength,page=5]{poles+first_contour.pdf}}%
    \put(0.31129839,0.36326571){\color[rgb]{0,0,0}\makebox(0,0)[lt]{\lineheight{1.25}\smash{\begin{tabular}[t]{l}$\mathcal{C}$\end{tabular}}}}%
  \end{picture}%
\endgroup%
 		\caption{Initial contour $\mathcal{C}$ from eq.~\eqref{eq:contour_int_C} enclosing the poles of the Fermi-Dirac function $\frac{1}{\eto{L_t z}+1}$.}\label{fig:contour_int_C}
	\end{figure}

	We employ the Matsubara technique
	\begin{align}
	C_0(t)&=\frac{1}{L_t}\sum_{p_0}\frac{1+n(m)\cos p_0}{m+\im \sin p_0}\eto{\im p_0 t}\\
	&=\frac{-1}{2\pi\im}\oint_\mathcal{C}\md z\, \frac{1}{\eto{L_t z}+1}\frac{1+n(m)\cos(-\im z)}{m+\im \sin (-iz)}\eto{z t}\\
	&=\frac{-1}{2\pi\im}\oint_\mathcal{C}\md z\, \frac{\eto{z t}}{\eto{L_t z}+1}\frac{1+n(m)\cosh z}{m+\sinh z}\label{eq:contour_int_C}
	\end{align}
	where the closed contour $\mathcal{C}$ has to be chosen such that it encloses the poles of the Fermi-Dirac function $\frac{1}{\eto{L_t z}+1}$ corresponding to the Matsubara frequencies $z=\im p_0=\im\frac{2\pi}{L_t}\left(j+\frac 12\right)$ with $j=0,\dots,L_t-1$ and no other poles, as visualised in figure~\ref{fig:contour_int_C}.
	
\begin{figure}[ht]
		\centering
\begingroup%
  \makeatletter%
  \providecommand\color[2][]{%
    \errmessage{(Inkscape) Color is used for the text in Inkscape, but the package 'color.sty' is not loaded}%
    \renewcommand\color[2][]{}%
  }%
  \providecommand\transparent[1]{%
    \errmessage{(Inkscape) Transparency is used (non-zero) for the text in Inkscape, but the package 'transparent.sty' is not loaded}%
    \renewcommand\transparent[1]{}%
  }%
  \providecommand\rotatebox[2]{#2}%
  \newcommand*\fsize{\dimexpr\f@size pt\relax}%
  \newcommand*\lineheight[1]{\fontsize{\fsize}{#1\fsize}\selectfont}%
  \ifx\svgwidth\undefined%
    \setlength{\unitlength}{474.00535059bp}%
    \ifx\svgscale\undefined%
      \relax%
    \else%
      \setlength{\unitlength}{\unitlength * \real{\svgscale}}%
    \fi%
  \else%
    \setlength{\unitlength}{\svgwidth}%
  \fi%
  \global\let\svgwidth\undefined%
  \global\let\svgscale\undefined%
  \makeatother%
  \begin{picture}(1,0.58547462)%
    \lineheight{1}%
    \setlength\tabcolsep{0pt}%
    \put(0,0){\includegraphics[width=\unitlength,page=1]{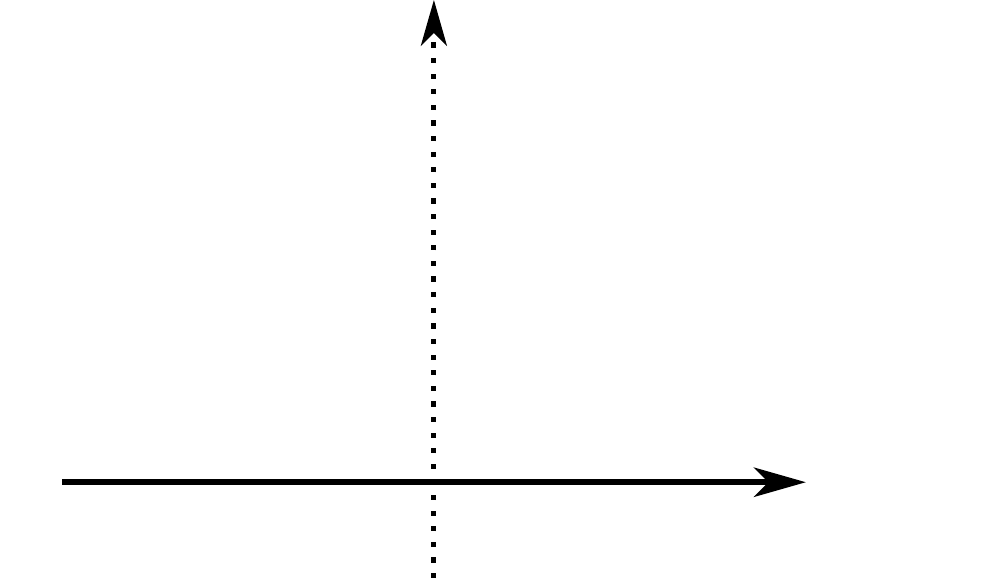}}%
    \put(0.76678209,0.04987154){\color[rgb]{0,0,0}\makebox(0,0)[lt]{\lineheight{1.25}\smash{\begin{tabular}[t]{l}$\real(z)$\end{tabular}}}}%
    \put(0.45197642,0.56349973){\color[rgb]{0,0,0}\makebox(0,0)[lt]{\lineheight{1.25}\smash{\begin{tabular}[t]{l}$\imag(z)$\end{tabular}}}}%
    \put(0,0){\includegraphics[width=\unitlength,page=2]{poles+deformed_contour.pdf}}%
    \put(0.15817421,0.06025631){\color[rgb]{0,0,0}\makebox(0,0)[lt]{\lineheight{1.25}\smash{\begin{tabular}[t]{l}$-\asinh m$\end{tabular}}}}%
    \put(0,0){\includegraphics[width=\unitlength,page=3]{poles+deformed_contour.pdf}}%
    \put(0.46213658,0.06929158){\color[rgb]{0,0,0}\makebox(0,0)[lt]{\lineheight{1.25}\smash{\begin{tabular}[t]{l}0\end{tabular}}}}%
    \put(0,0){\includegraphics[width=\unitlength,page=4]{poles+deformed_contour.pdf}}%
    \put(0.46995091,0.44848765){\color[rgb]{0,0,0}\makebox(0,0)[lt]{\lineheight{1.25}\smash{\begin{tabular}[t]{l}$\im 2\pi$\end{tabular}}}}%
    \put(0.45770522,0.17661863){\color[rgb]{0,0,0}\makebox(0,0)[lt]{\lineheight{1.25}\smash{\begin{tabular}[t]{l}$\im\pi/L_t$\end{tabular}}}}%
    \put(0.45770522,0.29622258){\color[rgb]{0,0,0}\makebox(0,0)[lt]{\lineheight{1.25}\smash{\begin{tabular}[t]{l}$3\im\pi/L_t$\end{tabular}}}}%
    \put(0,0){\includegraphics[width=\unitlength,page=5]{poles+deformed_contour.pdf}}%
    \put(0.59841625,0.14780284){\color[rgb]{0,0,0}\makebox(0,0)[lt]{\lineheight{1.25}\smash{\begin{tabular}[t]{l}$\mathcal{C}_1$\end{tabular}}}}%
    \put(0.23799405,0.40513311){\color[rgb]{0,0,0}\makebox(0,0)[lt]{\lineheight{1.25}\smash{\begin{tabular}[t]{l}$\mathcal{C}_3$\end{tabular}}}}%
    \put(0.01502777,0.22212493){\color[rgb]{0,0,0}\makebox(0,0)[lt]{\lineheight{1.25}\smash{\begin{tabular}[t]{l}$\mathcal{C}_4$\end{tabular}}}}%
    \put(0.81658757,0.32761433){\color[rgb]{0,0,0}\makebox(0,0)[lt]{\lineheight{1.25}\smash{\begin{tabular}[t]{l}$\mathcal{C}_2$\end{tabular}}}}%
  \end{picture}%
\endgroup%
 		\caption{Deformed contour $\mathcal{C}_1\cup\mathcal{C}_2\cup\mathcal{C}_3\cup\mathcal{C}_4$.}\label{fig:contour_deformed_C}
	\end{figure}

	The integrand is $2\pi \im$-periodic (reflecting the finite momenta range due to the lattice discretisation) and has singularities at $z=\im\frac{2\pi}{L_t}\left(\mathbb Z+\frac 12\right)$, $z=-\asinh m + \im 2\pi \mathbb Z$, and at $z=\asinh m + \im \pi + \im 2\pi \mathbb Z$. The two former are poles of first order whereas the latter can be lifted, i.e.\@ the numerator is zero as well, which corresponds to the disappearance of the doubler or back-propagating part (opposite real mass) and is exactly the reason we had to introduce the normalisation $n(m)$. Thus we can safely deform the contour $\mathcal{C}$ to the four paths shown in figure~\ref{fig:contour_deformed_C}
	\begin{align}
	\mathcal{C}_1&=\mathbb{R}+\im \varepsilon\,,\\
	\mathcal{C}_2&=\left[\infty+\im\varepsilon,\,\infty+2\pi\im-\im\varepsilon\right]\,,\\
	\mathcal{C}_3&=-\mathbb{R}+2\pi\im-\im \varepsilon\,,\\
	\mathcal{C}_4&=\left[-\infty+2\pi\im-\im\varepsilon,\,-\infty+\im\varepsilon\right]\,.
	\end{align}
	This means that we first integrate along the real axis shifted upwards by the infinitesimal imaginary parameter $\im\varepsilon$. Then at positive real infinity we go upwards to imaginary $2\pi\im-\im\varepsilon$. Next we go in negative direction parallel to the real axis. Finally we close the contour at negative real infinity going back down to the imaginary part $\im\varepsilon$.
	
	Let us consider $\mathcal{C}_2$ and $\mathcal{C}_4$ first. $t=0,\dots,L_t-1$, therefore at positive real infinity the integrand is exponentially suppressed by the Fermi-Dirac function, so $\mathcal{C}_2$ does not give any contribution. The integral along $\mathcal{C}_4$, in contrast, does not vanish for $t=0$. At negative real infinity the $\sinh$- and $\cosh$-terms dominate and the Fermi-Dirac function goes to one. So we get
	\begin{multline}
	\frac{-1}{2\pi\im}\int_{\mathcal{C}_4}\md z\, \frac{\eto{z t}}{\eto{L_t z}+1}\frac{1+n(m)\cosh z}{m+\sinh z}
	=\frac{-1}{2\pi\im}\int\limits_{-\infty+2\pi\im-\im\varepsilon}^{-\infty+\im\varepsilon}\md x\, \frac{\eto{t x}}{\eto{L_t x}+1}\frac{1+n(m)\cosh x}{m+\sinh x}\\
	=\frac{-1}{2\pi\im}\int\limits_{2\pi\im-\im\varepsilon}^{\im\varepsilon}\md x \left(-\delta_{t0}\right)
	=\frac{\delta_{t0}}{2\pi\im}\left(\im\varepsilon-\left(2\pi\im-\im\varepsilon\right)\right)
	=-\delta_{t0}
	\end{multline}
	for $\varepsilon\rightarrow 0$.
	
\begin{figure}[ht]
		\centering
\begingroup%
  \makeatletter%
  \providecommand\color[2][]{%
    \errmessage{(Inkscape) Color is used for the text in Inkscape, but the package 'color.sty' is not loaded}%
    \renewcommand\color[2][]{}%
  }%
  \providecommand\transparent[1]{%
    \errmessage{(Inkscape) Transparency is used (non-zero) for the text in Inkscape, but the package 'transparent.sty' is not loaded}%
    \renewcommand\transparent[1]{}%
  }%
  \providecommand\rotatebox[2]{#2}%
  \newcommand*\fsize{\dimexpr\f@size pt\relax}%
  \newcommand*\lineheight[1]{\fontsize{\fsize}{#1\fsize}\selectfont}%
  \ifx\svgwidth\undefined%
    \setlength{\unitlength}{412.5462367bp}%
    \ifx\svgscale\undefined%
      \relax%
    \else%
      \setlength{\unitlength}{\unitlength * \real{\svgscale}}%
    \fi%
  \else%
    \setlength{\unitlength}{\svgwidth}%
  \fi%
  \global\let\svgwidth\undefined%
  \global\let\svgscale\undefined%
  \makeatother%
  \begin{picture}(1,0.67269576)%
    \lineheight{1}%
    \setlength\tabcolsep{0pt}%
    \put(0,0){\includegraphics[width=\unitlength,page=1]{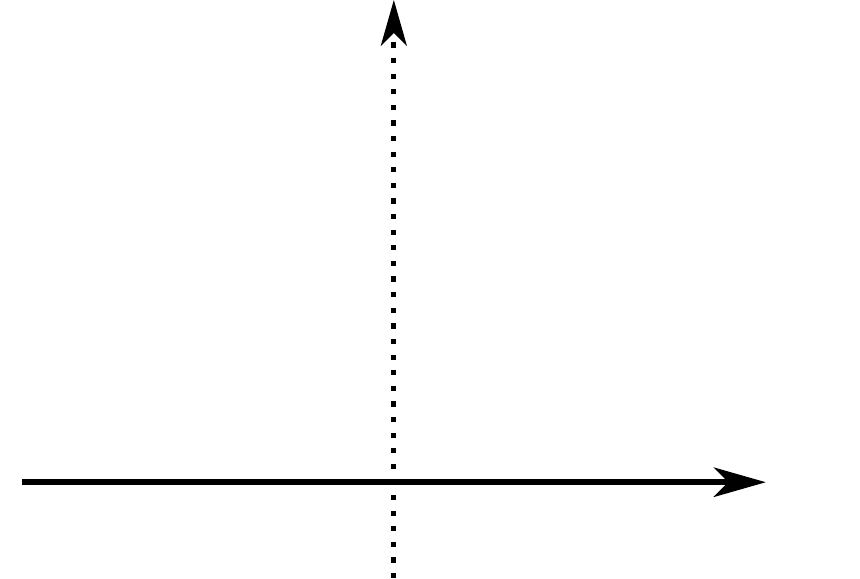}}%
    \put(0.88532299,0.12274832){\color[rgb]{0,0,0}\makebox(0,0)[lt]{\lineheight{1.25}\smash{\begin{tabular}[t]{l}$\real(z)$\end{tabular}}}}%
    \put(0.47271641,0.64744716){\color[rgb]{0,0,0}\makebox(0,0)[lt]{\lineheight{1.25}\smash{\begin{tabular}[t]{l}$\imag(z)$\end{tabular}}}}%
    \put(0,0){\includegraphics[width=\unitlength,page=2]{poles+shifted_contour.pdf}}%
    \put(0.20422822,0.08014082){\color[rgb]{0,0,0}\makebox(0,0)[lt]{\lineheight{1.25}\smash{\begin{tabular}[t]{l}$-\asinh m$\end{tabular}}}}%
    \put(0,0){\includegraphics[width=\unitlength,page=3]{poles+shifted_contour.pdf}}%
    \put(0.48439018,0.0796143){\color[rgb]{0,0,0}\makebox(0,0)[lt]{\lineheight{1.25}\smash{\begin{tabular}[t]{l}0\end{tabular}}}}%
    \put(0,0){\includegraphics[width=\unitlength,page=4]{poles+shifted_contour.pdf}}%
    \put(0.49336865,0.51530114){\color[rgb]{0,0,0}\makebox(0,0)[lt]{\lineheight{1.25}\smash{\begin{tabular}[t]{l}$\im 2\pi$\end{tabular}}}}%
    \put(0.47929866,0.20293041){\color[rgb]{0,0,0}\makebox(0,0)[lt]{\lineheight{1.25}\smash{\begin{tabular}[t]{l}$\im\pi/L_t$\end{tabular}}}}%
    \put(0.47929866,0.34035237){\color[rgb]{0,0,0}\makebox(0,0)[lt]{\lineheight{1.25}\smash{\begin{tabular}[t]{l}$3\im\pi/L_t$\end{tabular}}}}%
    \put(0,0){\includegraphics[width=\unitlength,page=5]{poles+shifted_contour.pdf}}%
    \put(0.68096774,0.16982178){\color[rgb]{0,0,0}\makebox(0,0)[lt]{\lineheight{1.25}\smash{\begin{tabular}[t]{l}$\mathcal{C}_1$\end{tabular}}}}%
    \put(0,0){\includegraphics[width=\unitlength,page=6]{poles+shifted_contour.pdf}}%
    \put(0.26685155,0.03644478){\color[rgb]{0,0,0}\makebox(0,0)[lt]{\lineheight{1.25}\smash{\begin{tabular}[t]{l}$\mathcal{C}_3-2\pi\im$\end{tabular}}}}%
    \put(0,0){\includegraphics[width=\unitlength,page=7]{poles+shifted_contour.pdf}}%
  \end{picture}%
\endgroup%
 		\caption{Alternative closed contour $\mathcal{C'}\supset\mathcal{C}_1 \cup \mathcal{C}_3-2\pi\im$.}\label{fig:contour_shifted_C}
	\end{figure}

	$\eto{N_t z}$ and $\sinh^2\frac z2$ are both $2\pi\im$-periodic. Thus integrating along $\mathcal{C}_3$ is identical to integrating along the real axis shifted by $-\im\varepsilon$. The union $\mathcal{C}_1 \cup \mathcal{C}_3-2\pi\im$ together with infinitesimal closing sequences at $\pm\infty$ is again a closed contour $\mathcal{C'}$ around the real axis winding once in negative direction (see fig.\ref{fig:contour_shifted_C}). The corresponding integral can be performed using the residuum theorem and plugging in the single real first order pole $z_0=-\asinh m$. We get
	\begin{align}
	\frac{-1}{2\pi\im}\oint_{\mathcal{C'}}\md z\, \frac{\eto{z t}}{\eto{L_t z}+1}\frac{1+n(m)\cosh z}{m+\sinh z}
	&=\res_{z_0}\frac{\eto{z t}}{\eto{L_t z}+1}\frac{1+n(m)\cosh z}{m+\sinh z}\\
	&=\lim_{z\rightarrow z_0}\frac{\eto{z t}}{\eto{L_t z}+1}\left(1+n(m)\cosh z\right)\frac{z-z_0}{m+\sinh z}\\
	&= \frac{\eto{-\asinh m t}}{\eto{-\asinh m L_t}+1}\frac{1+n(m)\cosh\asinh m}{\cosh\asinh m}\\
	&= \frac{\eto{-\tilde m t}}{\eto{-\tilde m L_t}+1}\frac{2}{\sqrt{1+m^2}}\,,
	\end{align}
	where $\tilde m\coloneqq \asinh m$.
	
	Now we have all the ingredients to evaluate equation~\eqref{eq:contour_int_C}. It yields
	\begin{align}
		C_0(t) &=
	\frac{-1}{2\pi\im}\oint_\mathcal{C}\md z\, \frac{\eto{z t}}{\eto{L_t z}+1}\frac{1+n(m)\cosh z}{m+\sinh z}\\
	&=\frac{-1}{2\pi\im}\left(\int_{\mathcal{C}_4}\md z+\oint_\mathcal{C'}\md z\right) \frac{\eto{z t}}{\eto{L_t z}+1}\frac{1+n(m)\cosh z}{m+\sinh z}\\
	&=\frac{\eto{-\tilde m t}}{\eto{-\tilde m L_t}+1}\frac{2}{\sqrt{1+m^2}}-\delta_{t0}\,.
	\end{align}

	\subsection{Case including branch cuts}\label{sec:c0_tilde_derivation}
\begin{figure}[ht]
		\centering
\begingroup%
  \makeatletter%
  \providecommand\color[2][]{%
    \errmessage{(Inkscape) Color is used for the text in Inkscape, but the package 'color.sty' is not loaded}%
    \renewcommand\color[2][]{}%
  }%
  \providecommand\transparent[1]{%
    \errmessage{(Inkscape) Transparency is used (non-zero) for the text in Inkscape, but the package 'transparent.sty' is not loaded}%
    \renewcommand\transparent[1]{}%
  }%
  \providecommand\rotatebox[2]{#2}%
  \newcommand*\fsize{\dimexpr\f@size pt\relax}%
  \newcommand*\lineheight[1]{\fontsize{\fsize}{#1\fsize}\selectfont}%
  \ifx\svgwidth\undefined%
    \setlength{\unitlength}{433.80650905bp}%
    \ifx\svgscale\undefined%
      \relax%
    \else%
      \setlength{\unitlength}{\unitlength * \real{\svgscale}}%
    \fi%
  \else%
    \setlength{\unitlength}{\svgwidth}%
  \fi%
  \global\let\svgwidth\undefined%
  \global\let\svgscale\undefined%
  \makeatother%
  \begin{picture}(1,0.63972784)%
    \lineheight{1}%
    \setlength\tabcolsep{0pt}%
    \put(0,0){\includegraphics[width=\unitlength,page=1]{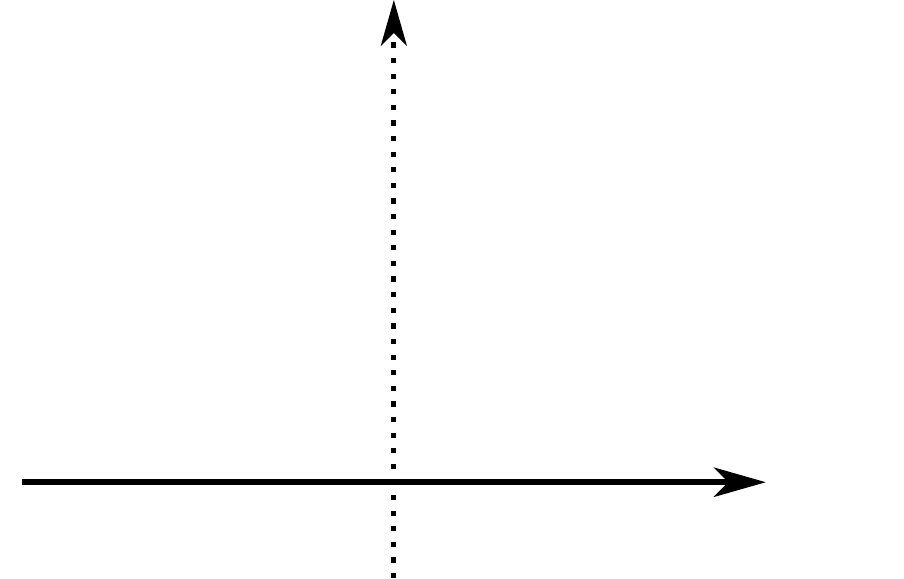}}%
    \put(0.8419345,0.11673259){\color[rgb]{0,0,0}\makebox(0,0)[lt]{\lineheight{1.25}\smash{\begin{tabular}[t]{l}$\real(z)$\end{tabular}}}}%
    \put(0.44954922,0.61571665){\color[rgb]{0,0,0}\makebox(0,0)[lt]{\lineheight{1.25}\smash{\begin{tabular}[t]{l}$\imag(z)$\end{tabular}}}}%
    \put(0,0){\includegraphics[width=\unitlength,page=2]{poles+branch_cuts.pdf}}%
    \put(0.28066335,0.07621323){\color[rgb]{0,0,0}\makebox(0,0)[lt]{\lineheight{1.25}\smash{\begin{tabular}[t]{l}$-\asinh m$\end{tabular}}}}%
    \put(0,0){\includegraphics[width=\unitlength,page=3]{poles+branch_cuts.pdf}}%
    \put(0.46065087,0.07571251){\color[rgb]{0,0,0}\makebox(0,0)[lt]{\lineheight{1.25}\smash{\begin{tabular}[t]{l}0\end{tabular}}}}%
    \put(0,0){\includegraphics[width=\unitlength,page=4]{poles+branch_cuts.pdf}}%
    \put(0.46918932,0.49004692){\color[rgb]{0,0,0}\makebox(0,0)[lt]{\lineheight{1.25}\smash{\begin{tabular}[t]{l}$\im 2\pi$\end{tabular}}}}%
    \put(0.45580888,0.19298506){\color[rgb]{0,0,0}\makebox(0,0)[lt]{\lineheight{1.25}\smash{\begin{tabular}[t]{l}$\im\pi/L_t$\end{tabular}}}}%
    \put(0.45580888,0.32367216){\color[rgb]{0,0,0}\makebox(0,0)[lt]{\lineheight{1.25}\smash{\begin{tabular}[t]{l}$3\im\pi/L_t$\end{tabular}}}}%
    \put(0,0){\includegraphics[width=\unitlength,page=5]{poles+branch_cuts.pdf}}%
    \put(0.57424421,0.16149905){\color[rgb]{0,0,0}\makebox(0,0)[lt]{\lineheight{1.25}\smash{\begin{tabular}[t]{l}$\mathcal{C}^0_1$\end{tabular}}}}%
    \put(0,0){\includegraphics[width=\unitlength,page=6]{poles+branch_cuts.pdf}}%
    \put(0.30218224,0.03465867){\color[rgb]{0,0,0}\makebox(0,0)[lt]{\lineheight{1.25}\smash{\begin{tabular}[t]{l}$\mathcal{C}^0_3$\end{tabular}}}}%
    \put(0,0){\includegraphics[width=\unitlength,page=7]{poles+branch_cuts.pdf}}%
    \put(0.76096189,0.16149905){\color[rgb]{0,0,0}\makebox(0,0)[lt]{\lineheight{1.25}\smash{\begin{tabular}[t]{l}$\mathcal{C}^+_1$\end{tabular}}}}%
    \put(0,0){\includegraphics[width=\unitlength,page=8]{poles+branch_cuts.pdf}}%
    \put(0.08669914,0.16149905){\color[rgb]{0,0,0}\makebox(0,0)[lt]{\lineheight{1.25}\smash{\begin{tabular}[t]{l}$\mathcal{C}^-_1$\end{tabular}}}}%
    \put(0.11200511,0.03465867){\color[rgb]{0,0,0}\makebox(0,0)[lt]{\lineheight{1.25}\smash{\begin{tabular}[t]{l}$\mathcal{C}^-_3$\end{tabular}}}}%
    \put(0.75860582,0.03465867){\color[rgb]{0,0,0}\makebox(0,0)[lt]{\lineheight{1.25}\smash{\begin{tabular}[t]{l}$\mathcal{C}^+_3$\end{tabular}}}}%
    \put(0,0){\includegraphics[width=\unitlength,page=9]{poles+branch_cuts.pdf}}%
    \put(0.70627545,0.07299576){\color[rgb]{0,0,0}\makebox(0,0)[lt]{\lineheight{1.25}\smash{\begin{tabular}[t]{l}$\ln2$\end{tabular}}}}%
    \put(0.12450689,0.07685484){\color[rgb]{0,0,0}\makebox(0,0)[lt]{\lineheight{1.25}\smash{\begin{tabular}[t]{l}$-\ln2$\end{tabular}}}}%
  \end{picture}%
\endgroup%
 		\caption{Split contours~\cref{eq:contour_C-,eq:contour_C0,eq:contour_C+} accounting for the branch cuts.}\label{fig:contour_with_branch_cuts}
	\end{figure}

	By and large, the derivation of $C_h(t)$ proceeds in the same way as that of $C_0(t)$ until the integral over $\mathcal C'$ has to be solved. This part turns out to be trickier as the paths $\mathcal C_1$ and $\mathcal C_3-2\pi\im$ cannot be connected at $\pm\infty$ because of the aforementioned branch cuts on the real axis starting at $\pm\acosh\frac54=\pm\ln2$. Instead we have to split both paths into three parts each
	\begin{align}
		\mathcal C_{1,3}^- &= \left[-\infty\pm\im\varepsilon,\,-\ln2\pm\im\varepsilon\right]\,,\label{eq:contour_C-}\\
		\mathcal C_{1,3}^0 &= \left[-\ln2\pm\im\varepsilon,\,\ln2\pm\im\varepsilon\right]\,,\label{eq:contour_C0}\\
		\mathcal C_{1,3}^+ &= \left[\ln2\pm\im\varepsilon,\,\infty\pm\im\varepsilon\right]\label{eq:contour_C+}
	\end{align}
	and bridge the gaps between them with infinitesimal paths orthogonal to the real axis as in figure~\ref{fig:contour_with_branch_cuts}. Thus we are left with $\mathcal C_1^0\cup\mathcal C_3^0$ enclosing the pole at $z=-\asinh m$ and yielding the same contribution as the integral over $\mathcal C'$, as well as the two paths along the branch cuts \begin{align}
		\mathcal C_\text{bc}^\pm &\coloneqq \mathcal C_1^\pm\cup\mathcal C_3^\pm\,.
	\end{align}
	
	Taking into account the integration directions dictated by the paths, we obtain
	\begin{align}
		\frac{-1}{2\pi\im}\int_{\mathcal{C}_\text{bc}^\pm}\md z\, \frac{\eto{z t}}{\eto{L_t z}+1}C_h(-\im z)
		&= \frac1\pi\,\Im\int_{\pm\ln2}^{\pm\infty}\md x\, \frac{\eto{t x}}{\eto{L_t x}+1}C_h(-\im x)\\
		&= \pm\frac1\pi\,\Im\int_{\ln2}^{\infty}\md x\, \frac{\eto{\pm t x}}{\eto{\pm L_t x}+1}C_h(\mp\im x)\\
		&\approx \pm\frac1\pi\,\Im\int_{\ln2}^{\infty}\md x\, \frac{\eto{\pm t x}}{\eto{\pm L_t x}+1}\frac{\sqrt{5-4 \cosh x}}{m\pm\sinh x}\\
		&= \pm\frac1\pi\int_{\ln2}^{\infty}\md x\, \frac{\eto{\pm t x}}{\eto{\pm L_t x}+1}\frac{\sqrt{4 \cosh x-5}}{m\pm\sinh x}\,.
	\end{align}
	To the best of our knowledge this integral has no exact analytic solution, so we used an approximation again, this time taking the scaling near $\ln2$ and the asymptotic behaviour into account:
	\begin{align}
		\frac{-1}{2\pi\im}\int_{\mathcal{C}_\text{bc}^+}\md z\, \frac{\eto{z t}}{\eto{L_t z}+1}C_h(-\im z)
		&\approx \frac1\pi\int_{0}^{\infty}\md x\,\frac{2^{t-L_t}}{m+\frac{3}{4}}\sqrt{3x}\,\eto{-\left(L_t-t+\frac{3}{4}\right) x}\\
		&=\sqrt{\frac3{4\pi}}\frac{1}{m+\frac34}\frac{2^{t-L_t}}{\left(L_t-t+\frac34\right)^{3/2}}\,,\\
		\frac{-1}{2\pi\im}\int_{\mathcal{C}_\text{bc}^-}\md z\, \frac{\eto{z t}}{\eto{L_t z}+1}C_h(-\im z)
		&\approx -\frac1\pi\int_{0}^{\infty}\md x\,\frac{2^{-t}}{m-\frac{3}{4}}\sqrt{3x}\,\eto{-\left(t+\frac{3}{4}\right) x}\\
		&=\sqrt{\frac3{4\pi}}\frac{1}{\frac34-m}\frac{2^{-t}}{\left(t+\frac34\right)^{3/2}}\,.
	\end{align}
\section{Algorithmic parameters $M$ and $L_s$}\label{sec:dw_height_separation}
\subsection{Influence of the Domain wall height $M$}
	The domain wall height $M$ has a significant impact on the propagators. 
For reasons of space 
we withhold the complete formula analogous to
equation~\eqref{eq:exact_mom_prop_h} and limit 
the discussion to numerical observations.
	
\begin{figure}[ht]
		\centering
		\includegraphics[width=.45\textwidth]{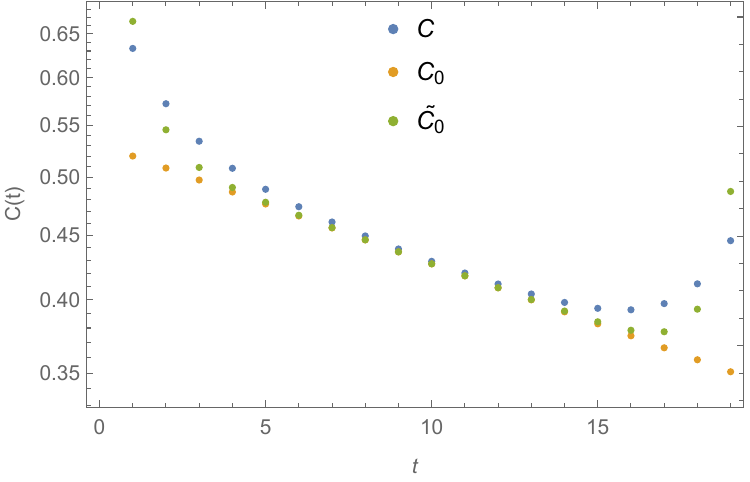}
		\includegraphics[width=.45\textwidth]{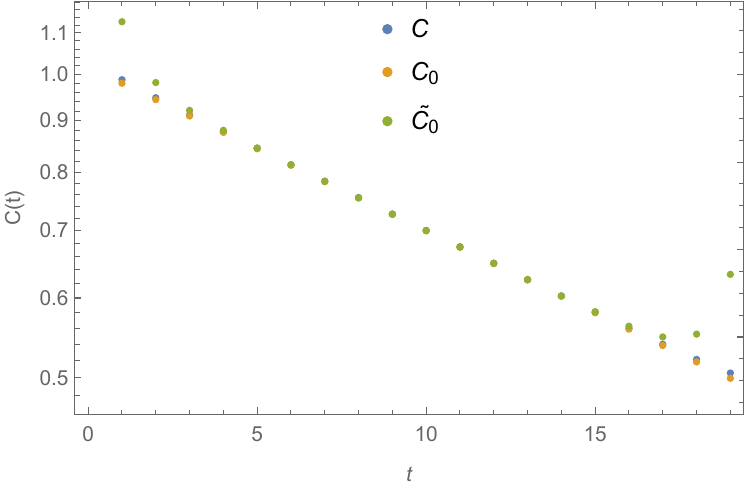}\\
		\includegraphics[width=.45\textwidth]{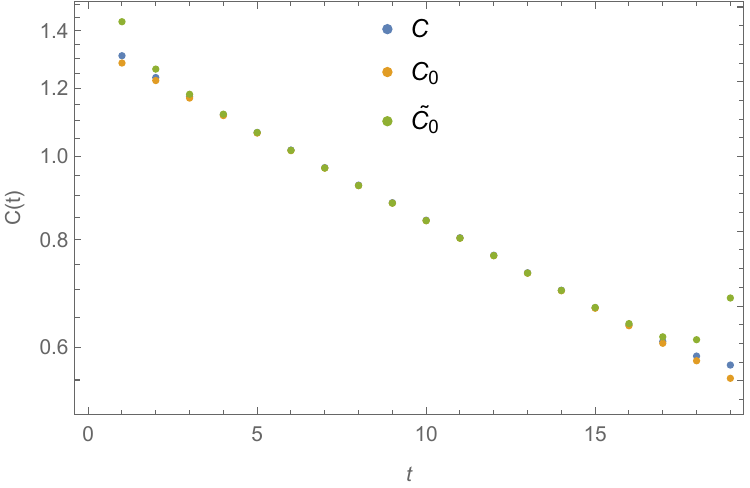}
		\includegraphics[width=.45\textwidth]{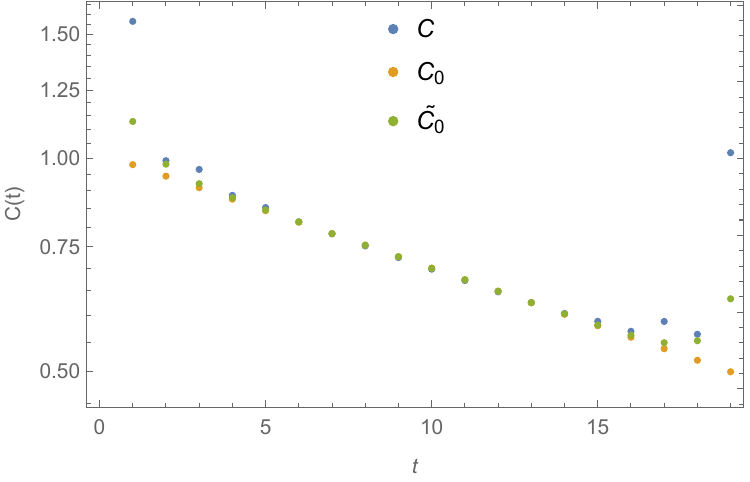}
		\caption{Exact and approximate free fermion propagators at zero spatial momentum in real space, bare mass $m=\num{0.05}$, $C_0$ and $\tilde C_0$ rescaled as in eq.~\eqref{eq:rescale_C0}, $m$ rescaled as in eq.~\eqref{eq:rescale_m}. Domain wall heights top: $M=\num{0.25}$, $M=\num{0.5}$; bottom: $M=\num{0.75}$, $M=\num{1.5}$.}\label{fig:free_prop_different_M_reweighted_m}
	\end{figure}

	In Fig.~\ref{fig:free_prop_different_M_reweighted_m}
	we show the time dependent
propagator for different domain wall heights where $C=C_h\equiv C_3$ is again calculated exactly.
	
	In order to approximate $C$, we have rescaled the approximate
correlators
	\begin{align}
		C_0(t) &\mapsto\left(1-(1-M)^2\right) C_0(t)\label{eq:rescale_C0}
	\end{align}
	and likewise for $\tilde C_0$, as well as the bare mass
	\begin{align}
	m &\mapsto m\left(1-(1-M)^2\right)\,.\label{eq:rescale_m}
	\end{align}
	We do not provide an analytic proof for the above formul\ae, though hard
staring at the propagator-like terms (22) in Ref.~\cite{Hands:2016foa} supports
their plausibility.
	With this additional modification $C$ is well approximated by $C_0$ even
for small domain wall heights $M\sim\num{0.25}$ 
as seen in Fig.~\ref{fig:free_prop_different_M_reweighted_m}.
The relative significance of the branch cut modelled by $\tilde C_0$, by contrast, is
sensitive to $M$.
$C$ shows rather unintuitive behaviour around the limits $t\approx 0$ and $t\approx L_t$. The deviations from the scaling at intermediate times, though always positive, appear to be smallest around $M\approx\num{0.5}$.
	
	\subsection{Influence of the Domain wall separation $L_s$}
	The finite-$L_s$ correction of $O(\eto{-\alpha L_s})$ in $C_h$, with
$\alpha(p)$ defined in (\ref{eq:alpha}), reads
	{%
\begin{align}
\begin{split}
			-& \left[ 4 \sin^2
{\textstyle\frac{p_0}{2}} \left(-3P +2 \cos
p_0 \left(P+2\right)-5\right)\right.\\
			&\quad\times \left(3 m^2 \left(P
+1\right)-\cos 2p_0 \left(m^2 \left(P+5\right)+3P
+13\right)+\cos p_0 \left(3 m^2+16P+47\right)\right.\\
			&\left.\left.+m \left(m \cos 3p_0-4 i \sin p_0
\left(\cos 2 p_0-2 \left(P+4\right) \cos p_0+3P
+8\right)\right)+\cos 3 p_0-13P-35\right)\right]\\
			&\left[\left(-2 \cos p_0+P+3\right)
(\cos p_0-1)\right.\\
			&\quad \left.\times \left(m^2P-2 \cos
p_0 \left(-m^2+P+3\right)-\left(m^2-1\right) \cos 2 p_0+2P
+5\right)^2\right]^{-1}
\end{split}
\end{align}}
with $P(p_0)\equiv\sqrt{5-4\cos p_0}$,
	and the corresponding term in $C_3$ reads
	{%
	\begin{align}
		\begin{split}
			& \left[ 4  \sin
^2{\textstyle\frac{p_0}{2}}\left(-3P +2 \cos p_0 \left(P+2\right)-5\right)\right.\\
			&\left.\times \left(-2 i \cos p_0 \left(2
m^2+P+2\right)+i \left(2 m^2 \left(P+3\right)+3P+5\right)+2 m \sin p_0 \left(-2
\cos p_0+P+3\right)\right)\right]\\
			&\left[P \left(-2 \cos p_0+P+3\right) (\cos p_0-1)\right.\\
			&\quad \left.\times \left(-m^2P+\left(m^2-1\right) \cos
2 p_0+2 \cos p_0 \left(-m^2+P+3\right)-2P -5\right)\right]^{-1}\,.
		\end{split}
	\end{align}}

\section{IR considerations}
\label{sec:IR}

The timeslice propagator measured in lattice simulations defined by
$C_{ft}(x_0)=\int d^2 x_\perp C_f(x_\perp,x_0)$ can be written
\begin{equation}
C_{ft}(x_0)={{\Gamma(2+\eta_\psi)}\over{2\eta_\psi x_0^{\eta_\psi}\Gamma(1-{{\eta_\psi}\over2})\Gamma(1+{{\eta_\psi}\over2})}}
=B(\eta_\psi)x_0^{-\eta_\psi}.
\label{eq:timeslice}
\end{equation}
When evaluating the timeslice correlator on a lattice of finite temporal
extent $L_t$, because of the algebraic decay of $C_f(x_0)$ it is necessary not
only to include the effects of a backwards
propagating signal, but also signals which have propagated $n$ times around
the lattice, ie.\ we need to incorporate ``image sources''. Each time a fermion
crosses the timelike boundary it picks up a minus sign due to boundary
conditions. The total is therefore
\begin{eqnarray}
C_{ft}(x_0;L)&=&{{B(\eta_\psi)}\over L_t^{\eta_\psi}}\sum_{n=0}^\infty \left[
{{(-1)^n}\over{\left({x_0\over L}+n\right)^{\eta_\psi}}}+
{{(-1)^n}\over{\left(1-{x_0\over L}+n\right)^{\eta_\psi}}}\right]\nonumber\\
&=&
{{B(\eta_\psi)}\over{(2L_t)^{\eta_\psi}}}\sum_{s={x_0\over L},1-{x_0\over L}}\left[
{\textstyle{\zeta(\eta_\psi,{s\over2})-\zeta(\eta_\psi,{1\over2}(1+s}))}
\right],\label{eq:hurwitz}
\end{eqnarray}
where $\zeta(\alpha,z)$ is the Hurwitz zeta function; 
the difference in
Eqn.~(\ref{eq:hurwitz}) is convergent for $\eta_\psi>0$ and given by an integral
suitable for numerical evaluation:
\begin{equation}
2^{-\eta_\psi}[{\textstyle{\zeta(\eta_\psi,{s\over2})-\zeta(\eta_\psi,{1\over2}(1+s))}}]=
{1\over{\Gamma(\eta_\psi)}}\int_0^\infty{{z^{\eta_\psi-1}e^{-sz}}\over{1+e^{-z}}}dz.
\end{equation}

\begin{figure}[h]
\begin{center}
  \includegraphics[width=0.9\columnwidth]{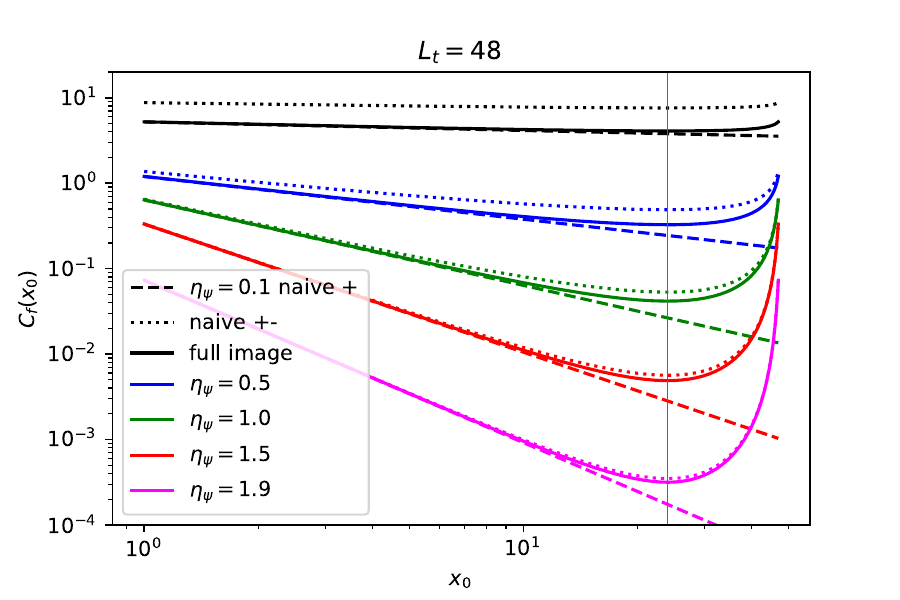}\\
  \caption{ $C_{ft}(x_0,L_t)$ for various $\eta_\psi$ with $L_t=48$.}
  \label{fig:correlator_IR}
\end{center}
\end{figure}
The resulting forms for $C_{ft}(x_0;L_t)$ are shown for various $\eta_\psi$ in
Fig.~\ref{fig:correlator_IR}. Dashed lines show the simple algebraic form
(\ref{eq:timeslice}), and dotted lines the result of a naive inclusion of just
a single backwards propagating signal, as done in conventional spectroscopy.
The antiperiodic boundary conditions significantly mitigate this
finite-$L_t$ artifact, particularly for small $\eta_\psi$, and indeed in making the signal
convergent in this limit. It is clear though that with $L_t=48$ it will be necessary to use a
formula such as (\ref{eq:hurwitz}) for precision fitting to $\eta_\psi$.
\printbibliography

\end{document}